\PassOptionsToPackage{usenames,dvipsnames}{xcolor}
\documentclass[twocolumn,twocolappendix]{aastex631}
\pdfoutput=1 %for arXiv submission
\usepackage[T1]{fontenc}
\usepackage{ae,aecompl}
\usepackage[utf8]{inputenc}
\usepackage[english]{babel}
\usepackage{amsmath,amstext}
\usepackage{apjfonts}
\usepackage{chngcntr}
\usepackage[figure,figure*]{hypcap}
\usepackage[hang]{footmisc}
\input{hyperlink-year-only-natbib-patch}

\newcommand*{\code}[1]{\ensuremath{\mathtt{#1}}}

\shorttitle{The DESI LOW-Z Secondary Target Sample}
\shortauthors{Darragh-Ford et al.}

\graphicspath{{./}{figures/}}

\begin{document}

\title{Target Selection and Sample Characterization for the DESI LOW-Z Secondary Target Program}

\author[0000-0002-8800-5652]{Elise Darragh-Ford}
\altaffiliation{edarragh@stanford.edu}
\affiliation{Kavli Institute for Particle Astrophysics and Cosmology and Department of Physics, Stanford University, Stanford, CA 94305, USA}
\affiliation{SLAC National Accelerator Laboratory, Menlo Park, CA 94025, USA}

\author[0000-0002-5077-881X]{John F. Wu}
\affiliation{Space Telescope Science Institute, 3700 San Martin Drive, Baltimore, MD 21218, USA}
\affiliation{Department of Physics \& Astronomy, Johns Hopkins University, 3400 North Charles Street, Baltimore, MD 21218, USA}

\author[0000-0002-1200-0820]{Yao-Yuan~Mao}
\affiliation{Department of Physics and Astronomy, University of Utah, Salt Lake City, UT 84112, USA}

\author[0000-0003-2229-011X]{Risa~H.~Wechsler}
\affiliation{Kavli Institute for Particle Astrophysics and Cosmology and Department of Physics, Stanford University, Stanford, CA 94305, USA}
\affiliation{SLAC National Accelerator Laboratory, Menlo Park, CA 94025, USA}

\author[0000-0002-7007-9725]{Marla~Geha}
\affiliation{Department of Astronomy, Yale University, New Haven, CT 06520, USA}

\author[0000-0002-2890-3725]{Jaime~E. ~Forero-Romero}\
\affil{Departamento de F\'{\i}sica, Universidad de los Andes, Cra. 1 No. 18A-10, Bogot\'{a}, Colombia}
\affil{Observatorio Astron\'omico, Universidad de los Andes, Cra. 1 No. 18A-10, Bogot\'{a}, Colombia}

\author[0000-0003-1197-0902]{ChangHoon Hahn}
\affil{Department of Astrophysical Sciences, Princeton University, Peyton Hall, Princeton NJ 08544, USA}
\affil{Lawrence Berkeley National Laboratory, 1 Cyclotron Road, Berkeley, CA 94720, USA}

\author[0000-0002-3204-1742]{Nitya Kallivayalil}
\affiliation{University of Virginia, Department of Astronomy, 530 McCormick Road, Charlottesville, VA 22904, USA}

\author[0000-0002-2733-4559]{John Moustakas} 
\affil{Department of Physics and Astronomy, Siena College, 515 Loudon Road, Loudonville, NY 12211, USA}

\author[0000-0002-1182-3825]{Ethan O.~Nadler}
\affiliation{Carnegie Observatories, 813 Santa Barbara Street, Pasadena, CA 91101, USA}
\affiliation{Department of Physics $\&$ Astronomy, University of Southern California, Los Angeles, CA, 90007, USA}

\author{Marta Nowotka}
\affiliation{Kavli Institute for Particle Astrophysics and Cosmology and Department of Physics, Stanford University, Stanford, CA 94305, USA}
\affiliation{SLAC National Accelerator Laboratory, Menlo Park, CA 94025, USA}

\author[0000-0003-4797-7030]{J.~E.~G.~Peek}
\affiliation{Space Telescope Science Institute, 3700 San Martin Drive, Baltimore, MD 21218, USA}
\affiliation{Department of Physics \& Astronomy,
Johns Hopkins University, 3400 North Charles Street, Baltimore, MD 21218, USA}

\author[0000-0002-9599-310X]{Erik~J.~Tollerud}
\affiliation{Space Telescope Science Institute, 3700 San Martin Drive, Baltimore, MD 21218, USA}

\author[0000-0001-6065-7483]{Benjamin~Weiner}
\affiliation{Department of Astronomy and Steward Observatory, University of Arizona, Tucson, AZ 85721, USA}

\author{J.~Aguilar}
\affil{Lawrence Berkeley National Laboratory, 1 Cyclotron Road, Berkeley, CA 94720, USA}

\author[0000-0001-6098-7247]{S.~Ahlen}
\affil{Physics Dept., Boston University, 590 Commonwealth Avenue, Boston, MA 02215, USA}

\author{D.~Brooks}
\affil{Department of Physics \& Astronomy, University College London, Gower Street, London, WC1E 6BT, UK}

\author[0000-0001-8274-158X]{A.P.~Cooper}
\affil{Institute of Astronomy and Department of Physics, National Tsing Hua University, 101 Kuang-Fu Rd. Sec. 2, Hsinchu 30013, Taiwan}

\author{A.~de la Macorra}
\affil{Instituto de F\'{\i}sica, Universidad Nacional Aut\'{o}noma de M\'{e}xico,  Cd. de M\'{e}xico  C.P. 04510,  M\'{e}xico}

\author[0000-0002-4928-4003]{A.~Dey}
\affil{NSF's NOIRLab, 950 N. Cherry Ave., Tucson, AZ 85719, USA}

\author{K.~Fanning}
\affil{Department of Physics, University of Michigan, Ann Arbor, MI 48109, USA}

\author[0000-0002-3033-7312]{A.~Font-Ribera}
\affil{Institut de F\'{i}sica d'Altes Energies (IFAE), The Barcelona Institute of Science and Technology, Campus UAB, 08193 Bellaterra Barcelona, Spain} 

\author[0000-0003-3142-233X]{S.~Gontcho A Gontcho}
\affil{Lawrence Berkeley National Laboratory, 1 Cyclotron Road, Berkeley, CA 94720, USA}

\author{K.~Honscheid}
\affil{Center for Cosmology and AstroParticle Physics, The Ohio State University, 191 West Woodruff Avenue, Columbus, OH 43210, USA}
\affil{Department of Physics, The Ohio State University, 191 West Woodruff Avenue, Columbus, OH 43210, USA}

\author[0000-0003-3510-7134]{T.~Kisner}
\affil{Lawrence Berkeley National Laboratory, 1 Cyclotron Road, Berkeley, CA 94720, USA} 

\author[0000-0001-6356-7424]{Anthony~Kremin}
\affil{Lawrence Berkeley National Laboratory, 1 Cyclotron Road, Berkeley, CA 94720, USA} 

\author[0000-0003-1838-8528]{M.~Landriau}
\affil{Lawrence Berkeley National Laboratory, 1 Cyclotron Road, Berkeley, CA 94720, USA} 

\author[0000-0003-1887-1018]{Michael E.~Levi}
\affil{Lawrence Berkeley National Laboratory, 1 Cyclotron Road, Berkeley, CA 94720, USA} 

\author[0000-0002-4279-4182]{P.~Martini}
\affil{Center for Cosmology and AstroParticle Physics, The Ohio State University, 191 West Woodruff Avenue, Columbus, OH 43210, USA}
\affil{Department of Astronomy, The Ohio State University, 4055 McPherson Laboratory, 140 W 18th Avenue, Columbus, OH 43210, USA}

\author[0000-0002-1125-7384]{Aaron M. Meisner}
\affil{NSF's NOIRLab, 950 N. Cherry Ave., Tucson, AZ 85719, USA} 

\author{R.~Miquel}
\affil{Instituci\'{o} Catalana de Recerca i Estudis Avan\c{c}ats, Passeig de Llu\'{\i}s Companys, 23, 08010 Barcelona, Spain}
\affil{Institut de F\'{i}sica d'Altes Energies (IFAE), The Barcelona Institute of Science and Technology, Campus UAB, 08193 Bellaterra Barcelona, Spain}

\author{Adam~D. ~Myers}
\affil{Department of Physics \& Astronomy, University  of Wyoming, 1000 E. University, Dept.~3905, Laramie, WY 82071, USA}

\author[0000-0001-6590-8122]{Jundan Nie}
\affil{National Astronomical Observatories, Chinese Academy of Sciences, A20 Datun Rd., Chaoyang District, Beijing, 100012, P.R. China} 

\author[0000-0003-3188-784X]{N.~Palanque-Delabrouille}
\affil{IRFU, CEA, Universit\'{e} Paris-Saclay, F-91191 Gif-sur-Yvette, France} 
\affil{Lawrence Berkeley National Laboratory, 1 Cyclotron Road, Berkeley, CA 94720, USA}

\author[0000-0002-0644-5727]{W.J.~Percival}
\affil{Department of Physics and Astronomy, University of Waterloo, 200 University Ave W, Waterloo, ON N2L 3G1, Canada"}
\affil{Perimeter Institute for Theoretical Physics, 31 Caroline St. North, Waterloo, ON N2L 2Y5, Canada}
\affil{Waterloo Centre for Astrophysics, University of Waterloo, 200 University Ave W, Waterloo, ON N2L 3G1, Canada}

\author[0000-0001-7145-8674]{F.~Prada}
\affil{Instituto de Astrof\'{i}sica de Andaluc\'{i}a (CSIC), Glorieta de la Astronom\'{i}a, s/n, E-18008 Granada, Spain}

\author{D.~Schlegel}
\affil{Lawrence Berkeley National Laboratory, 1 Cyclotron Road, Berkeley, CA 94720, USA}

\author{M.~Schubnell}
\affil{Department of Physics, University of Michigan, Ann Arbor, MI 48109, USA}
\affil{University of Michigan, Ann Arbor, MI 48109, USA}

\author[0000-0003-1704-0781]{Gregory~Tarl\'{e}}
\affil{University of Michigan, Ann Arbor, MI 48109, USA}

\author{M.~Vargas-Maga\~na}
\affil{Instituto de F\'{\i}sica, Universidad Nacional Aut\'{o}noma de M\'{e}xico,  Cd. de M\'{e}xico  C.P. 04510,  M\'{e}xico}

\author[0000-0002-4135-0977]{Zhimin~Zhou}
\affil{National Astronomical Observatories, Chinese Academy of Sciences, A20 Datun Rd., Chaoyang District, Beijing, 100012, P.R. China}

\author[0000-0002-6684-3997]{H.~Zou}
\affil{National Astronomical Observatories, Chinese Academy of Sciences, A20 Datun Rd., Chaoyang District, Beijing, 100012, P.R. China}

\begin{abstract}
We introduce the DESI LOW-Z Secondary Target Survey, which combines the wide-area capabilities of the Dark Energy Spectroscopic Instrument (DESI) with an efficient, low-redshift target selection method. Our selection consists of a set of color and surface brightness cuts, combined with modern machine learning methods, to target low-redshift dwarf galaxies ($z$ < 0.03) between $19 < r < 21$ with high completeness. We employ a convolutional neural network (CNN) to select high-priority targets. The LOW-Z survey has already obtained over 22,000 redshifts of dwarf galaxies (M$_* < 10^9$ M$_\odot$), comparable to the number of dwarf galaxies discovered in SDSS-DR8 and GAMA. As a spare fiber survey, LOW-Z currently receives fiber allocation for just $\sim 50\%$ of its targets. However, we estimate that our selection is highly complete: for galaxies at $z < 0.03$ within our magnitude limits, we achieve better than $95\%$ completeness with $\sim 1\%$ efficiency using catalog-level photometric cuts. We also demonstrate that our CNN selections $z<0.03$ galaxies from the photometric cuts subsample at least ten times more efficiently while maintaining high completeness. The full five-year DESI program will expand the LOW-Z sample, densely mapping the low-redshift Universe, providing an unprecedented sample of dwarf galaxies, and providing critical information about how to pursue effective and efficient low-redshift surveys.

%Our data validation shows that the CNN can achieve 90\% completeness at $z < 0.03$ with 20\% efficiency at selecting low-redshift galaxies, compared to compared to efficiencies of $\sim 1\%$ using traditional photometric methods. 
\end{abstract}

\keywords{\href{http://astrothesaurus.org/uat/416}{Redshift surveys (1378)}; \href{http://astrothesaurus.org/uat/1965}{Computational methods (1965)}; \href{http://astrothesaurus.org/uat/416}{Dwarf galaxies (416)}; \href{http://astrothesaurus.org/uat/940}{Low surface brightness galaxies (940)}}
\section{Introduction} \label{sec:intro}

Mapping the low-redshift Universe with a dense galaxy survey is a key goal of astronomy and cosmology, with diverse science applications, including understanding the properties of dwarf galaxies, identifying transient and gravitational wave hosts, measuring peculiar velocities, and mapping the detailed connection between galaxies and the matter density. 

A large sample of low-redshift dwarf galaxies can inform several important aspects of galaxy evolution, quasar physics, and dark matter physics. This includes studying the dwarf galaxy luminosity function, the best current estimates of which are from Sloan Digital Sky Survey \citep[SDSS,][]{Blanton_2005}, Galaxy And Mass Assembly Survey \citep[GAMA,][]{Loveday150501003}, and H\,{\sc i} measurements \citep[e.g.,][]{Jones2018}. However, for the faintest objects, large samples remain lacking. 
Measurements of the faintest end of the luminosity function and of the clustering properties of dwarf galaxies can help address important uncertainties in the galaxy--halo connection, such as what kind of halos do dwarf field galaxies live in and the efficiency of galaxy formation and baryonic feedback at these scales \citep{WechslerTinker}. Many of these faint galaxies also exist as satellites in larger halos, which allows us to estimate the scatter in the galaxy--halo connection through satellite kinematics \citep{Cao}. This scatter constrains the correlation between galaxy formation and halo formation, providing a critical test of galaxy formation models. Additionally, characterizing the field dwarf galaxy luminosity function serves as a stepping stone to place studies of ultra-faint dwarf galaxies in the Local Volume (e.g., \citealt{Martin_2016, Drlica_Wagner_2020, Nadler, Carlsten_2022,Nashimoto22}) in a cosmological context, reducing key uncertainties in these analyses, and connecting near-field studies to outstanding questions of halo and galaxy assembly bias.

Furthermore, obtaining larger samples of field dwarf galaxies can help reduce uncertainties on quenched fraction measurements at the faint end (e.g., \citealt{Geha_2012}), improving our understanding of low-mass galaxy formation. In addition, a wide-field sample of dwarf galaxies can be used to study the effects of environment on quenching \citep{Davies2019} and place constraints on galactic conformity \citep{Treyer2018}. A better understanding of quenching at low stellar masses can help determine the key feedback processes relevant for dwarf galaxies, for example, understanding where reionization vs. environmentally driven quenching dominates. 

A comprehensive catalog of low-redshift galaxies is also relevant to the task of efficiently identifying transient and gravitational wave hosts. For example, the upcoming Laser Interferometer Gravitational-Wave Observatory (LIGO) run expects to be sensitive to binary neutron star mergers out to 160--190 Mpc for the two original detectors, with the Virgo and KAGRA instruments having a more limited range \citep{ligo2020}. Despite the relatively small distances, optical follow-up is limited by relatively poor source localization---$10^2$--$10^3$ $\mathrm{deg}^2$ for two detectors, $10$--$10^2$ $\mathrm{deg}^2$ for three detectors, and $\lesssim 10\ \mathrm{deg}^2$ for four detectors---along with the sheer number density of all galaxies ($\sim 3000$ per $\mathrm{deg}^2$ at $r < 21$), only $\sim 10$ per square degree of which we expect to be truly low-redshift. Thus, a comprehensive catalog of low-redshift objects significantly reduces the number of potential host galaxies for a given event, increasing the likelihood of the successful observation of an optical counterpart. Such a catalog could also provide redshifts for standard siren measurements of the Hubble constant \citep{schutz,Abbott2017,Palmese2021, Chen_2022}.

The Dark Energy Spectroscopic Instrument (DESI) \citep{DESI2022} is an excellent tool for providing a large area, low-redshift spectroscopic survey. 
DESI, on the 4-meter Mayall telescope at Kitt Peak National Observatory, is a new massively multiplexed instrument capable of taking spectra of 5000 objects simultaneously, with a target density of $\sim 700$ objects per deg$^2$ and a spectral resolution of  $2000 <  \lambda/\Delta \lambda < 5500$ \citep{DESI2016, Silber2022, DESIcorrector2022}. The DESI Bright Galaxy Survey (BGS) is already set to enhance existing surveys by going significantly deeper than SDSS and wider than GAMA. Still, it is limited to $r < 19.5$ for its main magnitude-limited sample \citep{Hahn2022}. 

 Part of the difficulty in obtaining comprehensive samples of faint, low-redshift objects is due to the fact that although these objects are nearby on cosmic scales, separating them from the dominant background of high-redshift objects remains challenging. Significant effort has gone into accurate photometric redshifts (photo-$z$'s) for galaxy evolution and cosmology \citep{Baum1962, Benitez2000, Collister2004, Feldmann2006, Ilbert_2006, Brammer2008,Lee_2020,Li2022}, and also into designing photometric cuts to efficiently select high-redshift objects \citep[e.g.][]{Steidel1996,Daddi_2004, Finkelstein2015, Bouwens2015, Ono2018, Bowler_2020, Kauffmann2022}. However, analogous algorithms for selecting low-redshift objects have historically received less attention. This has made low-redshift surveys costly and time-consuming, as they must either accept a high rate of contamination of higher redshift objects or invest significant time into cleaning photometric catalogs by eye. 

Recent efforts to design photometric cuts for efficient low-redshift galaxy selection and more accurate low-redshift photo-$z$'s have produced impressive results. Machine learning methods have been able to achieve high accuracy in photo-$z$'s for the lowest redshift objects \citep{Pasquet2019, Dey_2022}. Meanwhile, the Satellites Around Galactic Analogs (SAGA) Survey \citep{Geha_2017, mao2020saga} has made significant progress towards converging on a set of photometric cuts optimized for low-redshift science targets that significantly reduce the target density while retaining high purity out to $z \sim 0.03$. This was validated using a targeted redshift survey of about 67,000 objects around very nearby (z < 0.01) galaxies. 

Here, we present the new DESI LOW-Z survey, designed to efficiently target faint, low-redshift ($z < 0.03$) objects. LOW-Z is a DESI spare fiber program, meaning it takes advantage of fibers not being used for primary DESI targets. The LOW-Z target selection strategy builds off of work done by the SAGA Survey in two key ways: (1) to define catalog-level cuts that efficiently select low-redshift galaxies, and (2) as a training set for a CNN that can increase this efficiency further using imaging data.  In this work, we detail the LOW-Z target selection strategy and characterize its {\it efficiency} (purity) and {\it completeness} at selecting $z < 0.03$ targets relative to a full magnitude limited survey. We do not attempt to account for surface brightness or other forms of incompleteness in the underlying photometric catalogs. In addition, while we characterize our redshift failure rates and fiber allocation fraction, correcting our completeness calculation for these effects requires careful modeling and is beyond the scope of this introductory paper. Since the main DESI survey strategy is optimized for cosmology, and as such, it is optimized to probe large volumes and measure the expansion history and growth rate of structure \citep{Levi2013}; at present, our program is a low-priority program using spare fibers. However, we show here that our strategy can already compete with previous low-redshift surveys and can extensively inform future programs for efficiently and completely surveying the low-redshift Universe.

\begin{deluxetable*}{l c c c c c c}
\centering
\tablecaption{\label{tab:tier_sum} Target density for the three tiers in the LOW-Z Survey. \emph{cols 1--2:} Submitted target densities for Y1 survey. \emph{cols 3--4:} Observed target densities for Y1 survey. All objects in cols 1-4 are between the Y1 LOW-Z magnitude cuts of $19 < r < 21$. \emph{cols 5--6:} Submitted target densities for Y2 survey. All objects are between the Y2 LOW-Z magnitude cuts of $19 < r < 21.15$.}
\tablehead{\colhead{Tier} & \colhead{Y1 Targets} & \colhead{Y1 Targets} & \colhead{Y1 Observed} & \colhead{Y1 Observed}& \colhead{Y2 Targets} & \colhead{Y2 Targets}
\\
 & All & BGS Overlap & All & BGS Overlap & All & BGS Overlap
}
\startdata
Tier 1 & $\phn{}22$ per deg$^2$ & $\phn{}\phn{}6$ per deg$^2$  &  $\phn{}11$ per deg$^2$ & $\phn{}\phn{}6$ per deg$^2$ &  $\phn{}97$ per deg$^2$ & $1.7$ per deg$^2$\\
Tier 2 & $\phn{}80$ per deg$^2$  & -- & \phn{}41 per deg$^2$ & -- &  $325$ per deg$^2$ & $1.3$ per deg$^2$\\
%Tier 3 & $200$ per deg$^2$ & $120$ per deg$^2$ & $150$ per deg$^2$ & $120$ per deg$^2$ &  -- & --\\
Tier 3A & $120$ per deg$^2$ & $120$ per deg$^2$ & $120$ per deg$^2$ & $120$ per deg$^2$ &  -- & --\\
Tier 3B & $\phn{}80$ per deg$^2$ & -- & $\phn{}30$ per deg$^2$ & -- &  -- & --\\
\hline
$z<0.03$ & -- & -- & $3.7$ per deg$^2$ & $1.6$ per deg$^2$ &  -- & --\\
\enddata 
\tablecomments{The full BGS target density is 1,400 targets per deg$^2$ (864 per deg$^2$ in the Bright sample and 533 per deg$^2$ in the Faint sample).}
\end{deluxetable*}

\begin{deluxetable*}{l c c c c}
\centering
\tablecaption{\label{tab:BGS_cuts} Color cuts for the BGS Bright and BGS Faint samples \citep{Hahn2022}}
\tablehead{\colhead{BGS Sample} & \colhead{$r$} & \colhead{$r_{\rm fib}$}  & \colhead{Color} & \colhead{Density}}
\startdata
BGS Bright & $r < 19.5$ & $r_{\rm fib} < 22.9$ & -- & 864 per deg$^2$ \\
BGS Faint & $19.5 < r < 20.175$  & $r_{\rm fib} < 21.5$ if color $\geq 0$ or  $r_{\rm fib} < 20.75$ & ($z - W1) - 1.2(g - r$) +1.2 & 533 per deg$^2$\\
\enddata
\end{deluxetable*}

\begin{figure*}
\centering
\includegraphics[width=\textwidth]{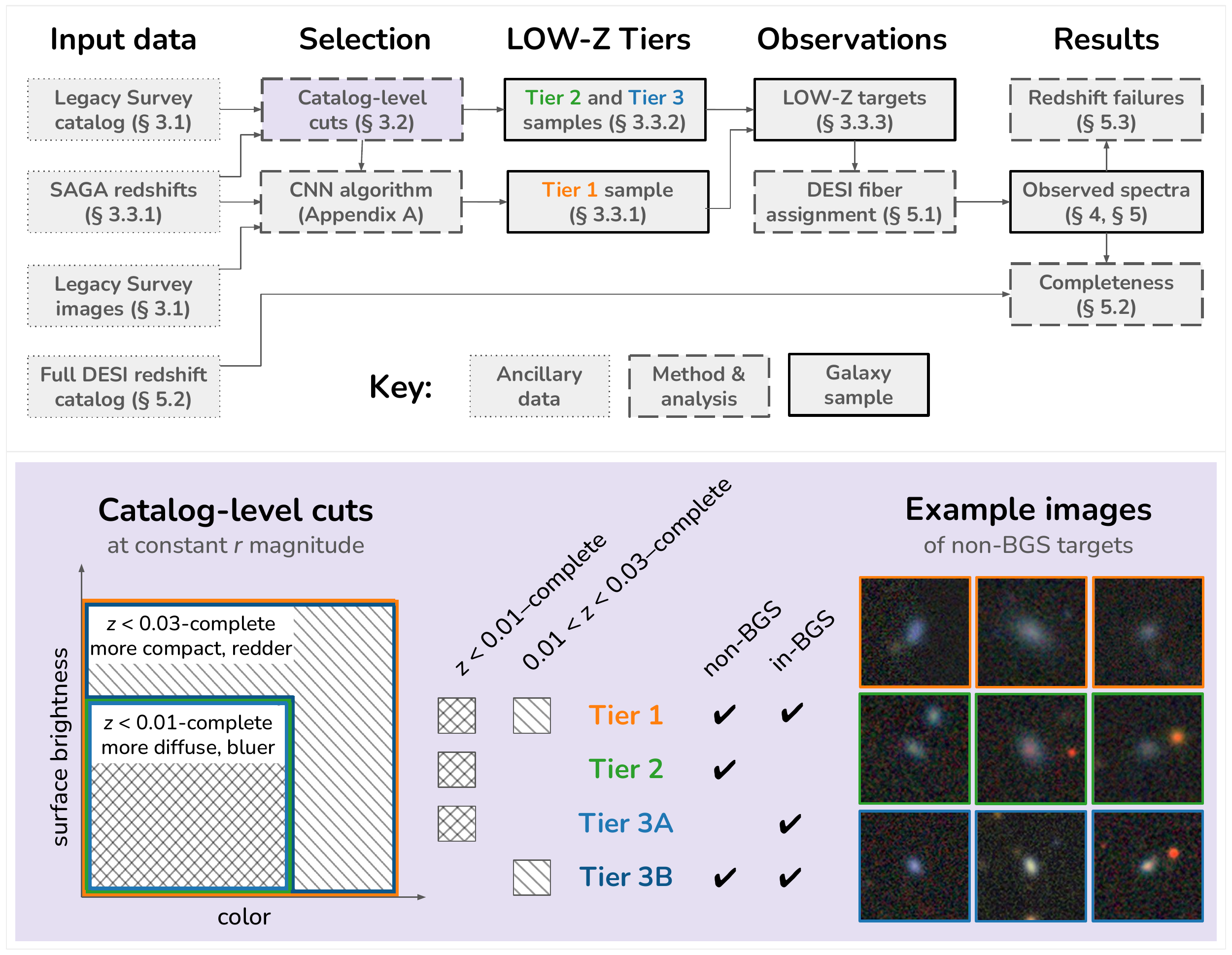}
\caption{(\textit{Upper}) A flow chart showing how targets are selected and observed in the DESI LOW-Z Survey. The chart references sections in the text (\S) where each process is described in more detail. (\textit{Lower left}) The photometric cuts for each LOW-Z tier are illustrated graphically using a color--surface brightness schematic diagram. (\textit{Lower center}) A table indicates overlap with BGS and targeting catalog surface density for each LOW-Z tier (see also Table~\ref{tab:tier_sum}). Tier 3 comprises two components, Tier 3A and Tier 3B, which have the same DESI fiber allocation priority (see Section~\ref{sec:tiers} for a description of the LOW-Z tiers). (\textit{Lower right}) For Tiers 1, 2, and 3, we display Legacy Survey DR9 $grz$-band $72^{\prime\prime} \times 72^{\prime\prime}$ image cutouts for three random non-BGS galaxy targets.}
% A flow chart showing the target selection for the DESI LOW-Z Survey. (1) Targets are selected from the Legacy Imaging Survey using photometric cuts. A sample of six randomly selected targets is shown as a visual example. (2) Targets are divided into three tiers using either a CNN (tier 1) or photometric criteria (tier 2 and tier 3). Visual examples of a target from each of the three tiers is shown next to the relevant label. (3) Targets are assigned to DESI spare fibers using the DESI fiber allocation algorithm. (4) Targets are observed by the DESI instrument, targets with successful redshifts determinations shown in the histogram to the right. Data from the SAGA survey is used both as training data for the CNN as well as to identify the photometric cuts used to target low-redshift galaxies.}
%Example selected  $z < 0.3$  objects from the DESI LOW-Z program, spanning the range of absolute $r$-band magnitude from each of the three tiers with the absolute $r-$magnitude and spectroscopic redshift of each object listed in the bottom right.}
\label{fig:flow_chart}
\end{figure*} 

\section{The LOW-Z Survey} 
The LOW-Z Survey is a DESI secondary target survey designed to target faint, low-redshift ($z < 0.03$) dwarf galaxies in dark time. DESI secondary target surveys make use of spare fibers (i.e., fibers that are not being used to target primary targets) to complement the main survey and its goals. LOW-Z targets are selected between $19 < r < 21$ using a set of color and surface brightness cuts. LOW-Z targets are further sorted into three tiers of priority, with the highest priority tier selected using a CNN trained on images of low-redshift galaxies from the SAGA survey. The remaining objects are split into two tiers based on their catalog-level photometric properties, with objects in the second tier corresponding to regions in parameter space where previous work indicates the majority of low-redshift dwarf galaxies are expected to lie \citep{Geha_2017, mao2020saga}. The full LOW-Z target sample consists of approximately 300 objects per square degree. However, the observed sample density is limited by the number of spare fibers available in a given pointing.  
In addition to getting redshifts for hundreds of thousands of low-redshift objects, the LOW-Z survey serves as a pilot program to refine methods for optimally selecting faint low-redshift targets for future campaigns in DESI-II\footnote{A potential extension of DESI that may run after the completion of the initial five-year survey.} and beyond. 

A flow chart describing the LOW-Z targeting strategy for the first year (Y1) of DESI operations can be seen in Figure \ref{fig:flow_chart}. This lays out the steps for target selection and tier identification for the LOW-Z program.  We discuss each of the steps individually in Section \ref{sec:selection}. We present the early LOW-Z sample in Section \ref{sec:main_results}, which consists of approximately 140,000 objects with spectra taken during the first year of DESI operations (approximately 17k of these objects were allocated fibers in dark time specifically as part of the LOW-Z program, while the remainder are objects that overlap with BGS and were allocated fibers in bright time). 
Using this sample as a benchmark, we validate the effectiveness of the LOW-Z Y1 targeting strategy for selecting a high completeness, magnitude-limited sample of low-redshift objects. Based on this analysis, we present slight modifications to the LOW-Z targeting strategy in Section \ref{sec:desiii} for the second year of DESI operations (Y2), which are currently ongoing. This includes using the Y1 data to provide updated target estimates from the CNN.

\subsection{LOW-Z as an extension of BGS}
%At the bright end, the LOW-Z sample overlaps with objects in the main BGS sample. 

The LOW-Z sample was designed specifically to complement the main DESI BGS sample. The DESI BGS consists of two samples: the BGS Bright sample, which targets all objects with $r < 19.5$, and the BGS Faint Sample, which targets objects between $19.5 < r < 20.175$ with an additional set of color-dependent cuts (Table \ref{tab:BGS_cuts}). 
The BGS program is significantly larger than the LOW-Z sample: 864 objects per deg$^2$ in the Bright sample and 533 objects per deg$^2$ in the Faint sample. %\footnote{$\sim 130$ per deg$^2$ of which overlap with the Year 1 LOW-Z targets.} 
It is expected to achieve $> 80\%$ fiber allocation for BGS Bright targets and $> 95\%$ redshift success rates for both samples. 

However, DESI BGS is a bright-time program, meaning that targets are observed during bright conditions (determined based on observing conditions such as seeing, transparency, airmass, and sky brightness). This means that BGS is limited in its ability to obtain redshifts for the faintest and lowest surface brightness objects. In contrast, LOW-Z is a dark-time program. This allows the LOW-Z survey to complement the BGS in two key ways: (1) LOW-Z goes over half a magnitude fainter than the BGS Faint sample and a full magnitude and a half fainter than the BGS Bright sample, helping to fill in objects at the faint end of the galaxy luminosity function; and (2) LOW-Z objects are observed in dark rather than bright time, which allows us to target objects fainter than the BGS fiber magnitude cut at $r_{\rm fib} = 22.9$ without drastically increasing our redshift failure rate (see discussion in Section \ref{sec:rshift_failure}), meaning that the LOW-Z sample will be more complete than BGS for very low surface brightness objects. 

Due to the DESI fiber assignment strategy, most objects that overlap between the two samples will be allocated fibers in bright time as part of the main BGS survey\footnote{We include these objects in the LOW-Z sample presented here, even though they were not formally targeted as part of the LOW-Z program.}. However, for BGS objects in the LOW-Z sample that do not receive fiber allocation in bright time, LOW-Z provides a second opportunity for fiber assignment with a higher likelihood of redshift success for objects with $22 < r_\mathrm{\rm fib} < 22.9$ (Section \ref{sec:rshift_failure}). Since BGS observations supersede LOW-Z observations in terms of priority, objects that overlap between the two samples and receive fiber allocation in bright time are removed from dark-time target lists.

\section{LOW-Z Y1 Targeting Strategy}
\label{sec:selection}

\subsection{Imaging Data}
\label{sec:cleaning_cuts}

We select objects using the catalog from the Data Release 9 (DR9) of the DESI Legacy Imaging Surveys \citep{Zou_2017, Dey2019_Legacy, dr9}\footnote{\url{https://www.legacysurvey.org/viewer/}}. The DR9 catalog consists of data from three imaging projects: The Beijing-Arizona Sky Survey (BASS), The DECam Legacy Survey (DECaLS), and The Mayall z-band Legacy Survey (MzLS). 

We use the \code{TYPE} flag to identify galaxies as all objects whose $\code{TYPE} \neq \code{PSF}$ and remove duplicated \textit{Gaia} entries using $\code{TYPE} \neq \code{DUP}$.
We use \code{FLUX} and \code{MW\_TRANSMISSION} to calculate dereddened magnitudes. We use \code{SHAPE\_R} as our effective photometric radius, $R_{r,\text{eff}}$. For bands in $\emph{grz}$ we additionally define \code{SIGMA\_GOOD} for the purpose of implementing quality cuts. Unless explicitly defined below, these quantities come directly from the DR9 catalog\footnote{\url{https://www.legacysurvey.org/dr9/catalogs/}} and the definitions can be found in the relevant citations above. 
\begin{equation*}
\code{SIGMA\_GOOD} = \left\{
\begin{array}{ll}
\code{FLUX} \times \sqrt{\code{FLUX\_IVAR}}, & \text{if } \code{RCHISQ} < 100;   \\
0, \text{otherwise}.
\end{array} \right.  
\end{equation*}

While the DR9 photometric catalog is generally very clean, it still contains some spurious objects, including shredded sources, false positive detections, and sources with highly overestimated magnitudes. We apply a set of quality cuts to remove the majority of these spurious objects from our targets. Specifically, we only include objects that satisfy all of the following criteria: 
\begin{align*}
&\code{SIGMA\_GOOD} \geq 5.0 \text{ (any two bands)}; \\
&\code{FRACFLUX} \leq 0.35 \text{ (any two bands)}; \\
&\code{RCHISQ} \leq 2.0 \text{ (any two bands)}; \\
&\code{SIGMA\_GOOD} \geq 30 \text{ or } \code{RCHISQ} \leq 0.85 \text{ (any two bands)}; \\
&\code{g} - \code{r} > -0.1. 
\end{align*}
These criteria were first developed for the SAGA Survey \citep[][]{mao2020saga}, and later adopted for cleaning the LOW-Z sample. The criteria on \code{SIGMA\_GOOD} aim to remove false positive detections, those on \code{FRACFLUX} aim to remove shredded sources from a brighter companion, those on \code{RCHISQ} aim to remove sources with very inaccurate model fits, and finally, those on $g-r$ aim to remove sources with very different fits in $g$ and $r$ bands. We visually inspect the resulting targets to set the thresholds in these criteria so that they remove the majority of these spurious objects without impacting our target selection completeness.

We exclude objects that are within 1.5 times the radius of an object in the Siena Galaxy Atlas (SGA) catalog \citep{sga} or within 4 times the half-light radius of any non-SGA objects in DR9 catalogs brighter than $r = 16$. Galactic radii in SGA are defined as the radius at the 25 mag arcsec$^{-2}$ surface brightness isophote. This was done as a further cleaning step to avoid targeting misidentified remnants of bright galaxies and was designed to remove only those objects that significantly overlap with the light of a brighter galaxy. This should not strongly impact the sample satellite galaxies in the LOW-Z survey. However, for a detailed comparison of the differential impact of environment on isolated and satellite dwarf galaxies, the LOW-Z sample is well-suited to comparison with satellites from the SAGA Survey, as the two were selected using nearly identical color and surface brightness criteria and span a similar range in magnitudes and distances.

\subsection{LOW-Z Catalog-Level Photometric Cuts}
\label{sec:photo_cuts}

Accurately identifying low-redshift galaxies using only photometric data is difficult, even when spectroscopic training sets are available. Most current photometric redshift algorithms have been trained on data that has been explicitly color selected for high-redshift galaxies. In addition,  low-redshift objects are vastly outnumbered by higher redshift objects in almost every available training set. There are a few thousand objects per $\mathrm{deg}^2$ between $19 < r < 21$, all but tens of which we expect to be bright galaxies at a higher redshift ($z > 0.03$). Thus, efficiently selecting low-redshift objects in this regime requires careful study.

Here, we present a set of catalog-level photometric cuts designed specifically for target selection of low-redshift ($z < 0.03$) objects to high completeness (hereafter referred to as the $z<0.03$-complete photometric cuts). These cuts are developed based on the photometric cuts first introduced by the SAGA Survey \citep{mao2020saga}. The SAGA Survey Stage II targeting cuts were tested extensively by the SAGA Survey team, including tests with a complete spectroscopic survey of objects around two SAGA systems. These cuts were found to be complete out to $z < 0.01$ at a target density of $\sim$ 200 objects per $\mathrm{deg}^2$; hence, we will refer to the SAGA Survey Stage II targeting cuts as the $z<0.01$-complete photometric cuts hereafter. The $z<0.03$-complete cuts presented here are identical to the $z<0.01$-complete photometric cuts but with an increase in the surface brightness and color thresholds used:

\begin{subequations}
\begin{align}
\mu_{r{_o}, \textrm{eff}} + \sigma_{\mu} - 0.7 \, (r_o - 14) &>
    \begin{cases}
        16.8  & (z<0.03{\rm -complete}) \\
        18.5  & (z<0.01{\rm -complete})
    \end{cases}
\label{eq:targeting-cuts-sb-r} \\
(g-r)_o - \sigma_{gr} + 0.06\,(r_o - 14) &<     \begin{cases}
        0.99  & (z<0.03{\rm -complete})\\
        0.9  & (z<0.01{\rm -complete})
    \end{cases} \label{eq:targeting-cuts-gr-r}
\end{align}%
\end{subequations}
where $r_o$, $g_o$ are the extinction-corrected $r$ and $g$-band apparent magnitudes respectively, $\mu_{r{_o}, \textrm{eff}}$ is effective surface brightness, $\sigma_{\mu}$ is the error on $\mu_{r{_o}, \textrm{eff}}$, and $\sigma_{gr} \equiv \sqrt{\sigma_g^2 + \sigma_r^2}$ is the error on the $(g-r)_o$ color. We calculate $\mu_{r{_o}, \textrm{eff}}$ and $\sigma_{\mu}$ analogously to \cite{mao2020saga}.  
We present the validation of the completeness $z<0.03$-complete photometric cuts in Section \ref{sec:selection_eff}. 

\subsection{LOW-Z Tier Assignment} \label{sec:tiers}
The full LOW-Z target sample consists of all objects at $19 < r < 21$ passing the $z<0.03$-complete photometric cuts (Equations \ref{eq:targeting-cuts-sb-r};\ref{eq:targeting-cuts-gr-r}). However, in order to maximize our observed sample of low-redshift objects, we split the target sample into three tiers, which roughly correspond to our expectation of a given object being legitimately low-redshift. A CNN algorithm selects the first tier, while the second and third tiers correspond to different regions in color--surface brightness parameter space. The LOW-Z tiers are hierarchical, such that objects in Tier 1 are excluded from Tier 2, and objects from Tiers 1 and 2 are excluded from Tier 3.

\subsubsection{Tier 1: CNN Selection from Imaging}
\label{sec:CNN_Y1}
We use a convolutional neural network (CNN) to select our {\bf Tier 1} sample on the basis of their imaging \citep{Wu2021xsaga}.
A CNN is a parametric model that can be optimized to make predictions purely from images as inputs. Understanding how CNNs work (as well as they do) remains an active field of research, but we attempt to provide some intuition here. A CNN can be thought of as a multi-scale matched-filtering algorithm with fully learnable filters \citep[see, e.g.,][]{Mallat2016}. In other words, the input image is decomposed into multi-color morphological features at various scales. Crucially, each convolution with a learned filter is also followed by a non-linear operation and a pooling layer, which decreases the resolution while increasing the receptive field. Additionally, residual layers in the CNN permit interactions between different scales \citep{He2016}. These ingredients enable the CNN developed here to efficiently identify low surface brightness features and other distinguishing elements of low-redshift galaxy images.

We trained a CNN to separate low-redshift ($z < 0.03$) galaxies from high-redshift ($z > 0.03$) using {\it grz}-band $144 \times 144$-pixel image cutouts from the DESI Legacy Imaging Surveys DR9. The CNN prediction $p_{\rm CNN}$ can range between 0 to 1, where 1 represents the highest confidence that the input image is a low-redshift system. CNN training details are presented in Appendix~\ref{sec:cnn-details}.
In the interest of incorporating all of the valuable data for training the CNN, we use the SAGA redshift catalog that is identical to the one used in \cite{Wu2021xsaga}. This catalog contains 112,016 galaxy redshifts that the SAGA Survey team has measured or compiled around SAGA hosts. Among these galaxy redshifts, 2,550 are at $z < 0.03$.
The majority (89\%) of the $z < 0.03$ galaxies in this catalog lie within the $z<0.01$-complete photometric cuts, and almost all (98.5\%) of the $z < 0.03$ galaxies lie within the $z<0.03$-complete photometric cuts.
Additional details about our spectroscopic data set can be found in Section~2.2 of \citet{Wu2021xsaga}.
%In Section~\ref{sec:cnn-retraining}, we apply an updated set of photometric cuts, retrain the CNN model using a combined SAGA and DESI redshift catalog, and propose new targets for DESI Year 2 and onward.

%\section{LOW-Z One-Percent and Year 1 Sample Selection}
%\label{sec:targeting}

%Our target sample consists of all objects between $19 < r < 21$ passing the $z<0.03$-complete photometric cuts (Equations \ref{eq:targeting-cuts-sb-r};\ref{eq:targeting-cuts-gr-r}). 
%The full sample of LOW-Z targets are then divided in three tiers of priority:

We use the CNN to select approximately 20 objects per square degree from the sample selected using $z<0.03$-complete photometric cuts. In other words, we train the CNN on the full SAGA redshift sample, including objects \textit{outside} the catalog level cuts, but we use the CNN to select targets from \textit{within} these cuts. In the north, we remove all objects with $p_{CNN} < 0.2503$, and in the south, those with $p_{CNN} < 0.3308$. We use slightly different cutoff thresholds in the two regions to ensure approximately constant density across the whole sky \citep[similar to the photometric offsets found in][]{Zarrouk2022}. 
From training and cross-validation experiments, we find that our CNN selection achieves $\sim 45\%$ purity and $\sim 85\%$ completeness on the SAGA redshift catalog.

\subsubsection{Tier 2 and Tier 3: Catalog-Level Photometric Selection}

{\bf Tier 2} and {\bf Tier 3} are selected using purely catalog-level photometric criteria. Tier 2 corresponds to objects within the $z<0.01$-complete photometric cuts outside of the BGS sample, while Tier 3 consists of objects in the $z<0.01$-complete photometric cuts that overlap with BGS (Tier 3A) as well as a random sampling of the remaining objects between the $z<0.01$-complete photometric cuts and the $z<0.03$-complete photometric cuts (Tier 3B). The objects in Tier 3B are, by definition, redder and more compact than the Tier 2 objects and thus have a lower probability of being legitimate low-redshift objects. 

In practice, the deprioritization of objects in the $z<0.01$-complete photometric cuts that overlap with BGS from Tier 2 to Tier 3A has a negligible impact on the number of these targets that are allocated fibers. This is due to the fact that BGS targets supersede LOW-Z targets in terms of fiber allocation priority, meaning that most of these objects will be assigned as bright-time targets as part of BGS and therefore will not be included as LOW-Z targets during fiber assignment. However, the split is useful for analysis as it ensures that all Tier 2 targets were specifically targeted as part of the LOW-Z program in dark time.  After removing the overlap with BGS, Tier 2 consists of approximately 80 objects per square degree. 

Finally, due to survey limitations for our total target density, we downsample objects in Tier 3B to approximately 80 objects per square degree. %This downsampling is only done for objects that do not overlap with the BGS main survey, as BGS objects are automatically included in targeting lists and thus do not affect the overall target density of the survey. 
The downsampled objects in Tier 3B are selected in the following two stages: (\textit{i}) all objects within the $z<0.01$-complete surface brightness cuts and between the $z<0.01$-complete and $z<0.03$-complete $g - r$ cuts ($\sim 40$ objects per deg$^2$); (\textit{ii}) random sample of the remaining objects between the $z<0.01$-complete and $z<0.03$-complete cuts, that are not in $(i)$ ($\sim 40$ objects per deg$^2$). We prioritize the redder objects in Tier 3B to ensure we have an accurate representation of the population of quenched low-mass dwarfs.  The random sampling in (\textit{ii}) downsamples the total number of objects by a factor of $\sim 2$ in these photometric regions. After downsampling, the total combined target density of Tiers 3A and 3B is approximately 200 objects per square degree.

\subsubsection{Summary of LOW-Z Tiers}  \label{sec:summary}
\textbf{Tier 1} ($\sim 22$ objects per deg$^2$) consists of objects selected by the CNN from the $z<0.03$-complete photometric cuts sample. Approximately six objects per square degree in this sample overlap with the DESI BGS sample. 

\textbf{Tier 2} ($\sim 80$ objects per deg$^2$) consists of objects from the $z<0.01$-complete photometric cuts sample that are outside of the main BGS color cuts (Table \ref{tab:BGS_cuts}) \citep{Hahn2022}. 

\textbf{Tier 3} ($\sim 200$ objects per deg$^2$) consists of Tier 3A --- the remaining objects from $z<0.01$-complete photometric cuts sample that overlap with the main BGS sample ($\sim 120$ objects per deg$^2$), as well as Tier 3B --- the objects from the $z<0.03$-complete photometric cuts sample that are outside of the $z<0.01$-complete photometric cuts ($\sim 80$ objects per deg$^2$). 

The density for each of the three tiers can be found in Table \ref{tab:tier_sum}. 
% Randomly selected objects from each of the three tiers can be seen in Figure \ref{fig:flow_chart}. 
On average, the CNN tends to select objects that are larger and have lower surface brightness. In addition, it selects a higher fraction of blue objects than the $z<0.01$-complete photometric cuts sample. Meanwhile, the $z<0.03$-complete photometric cuts sample is on average redder and more compact due to the relaxation of the $g - r$ and $\mu_{r, {\rm eff}}$ cuts (Equations \ref{eq:targeting-cuts-gr-r}; \ref{eq:targeting-cuts-sb-r}). 

Due to an error in target selection, Tier 1 and Tier 3 had slightly different selection criteria for the first few months of DESI year one (the sample considered here). Outside of the BGS color--magnitude cuts, Tier 1 and Tier 3B only contain objects in the $z<0.03$-complete photometric cuts region that are outside of both the $z<0.01$-complete surface brightness and $g - r$ cuts. This means that only objects with both $g - r$ or $\mu_{r, {\rm eff}}$ outside of the $z<0.01$-complete photometric cuts are included in the extended sample in dark time. In addition, in the northern sky, Tier 1 only contains objects within the $z<0.01$-complete photometric cuts.

\begin{deluxetable}{l c c c}
\centering
\tablecaption{\label{tab:dens_obs} Observed number of targets in LOW-Z survey split by tier and redshift.}
\tablehead{\colhead{} & \colhead{z < 0.01} & \colhead{z < 0.03} & \colhead{All redshifts}}
\startdata
One-Percent Survey \\ 
\hline
Tier 1 & 26 &  382 & \phn{}2,015\\
Tier 1 (excl. BGS) & 12 &  167 & \phantom{00,}992\\
Tier 2 & \phn{}5 &  100 & \phn{}7,445\\
Tier 3 & \phn{}3 &  179 & 27,021\\
Tier 3 (excl. BGS) & \phn{}2 &  \phn{}37 & \phn{}5,906\\
\hline
Main Survey \\ 
\hline
Tier 1 & 53 &  875 & \phn{}\phn{}4,618\\
Tier 1 (excl. BGS) & \phn{}4 &  \phn{}34 & \phantom{000,}163\\
Tier 2 & \phn{}1 &  \phn{}22 & \phn{}\phn{}2,034\\
Tier 3 & \phn{}3 &  461 & 100,353\\
Tier 3 (excl. BGS) & \phn{}0 &  \phn{}\phn{}4 & \phn{}\phn{}1,413\\
\enddata
\tablecomments{The One-Percent survey and main survey samples are non-overlapping. The Main Survey results presented here include just the first two months of DESI Y1.}
\end{deluxetable}

\begin{figure*}[htb!]
\centering
\includegraphics[width=0.74\columnwidth]{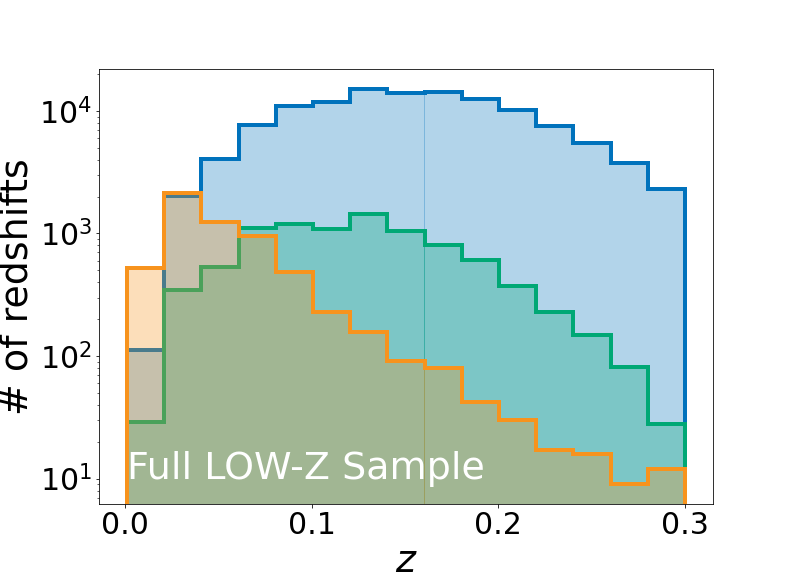}
\includegraphics[width=0.68\columnwidth]{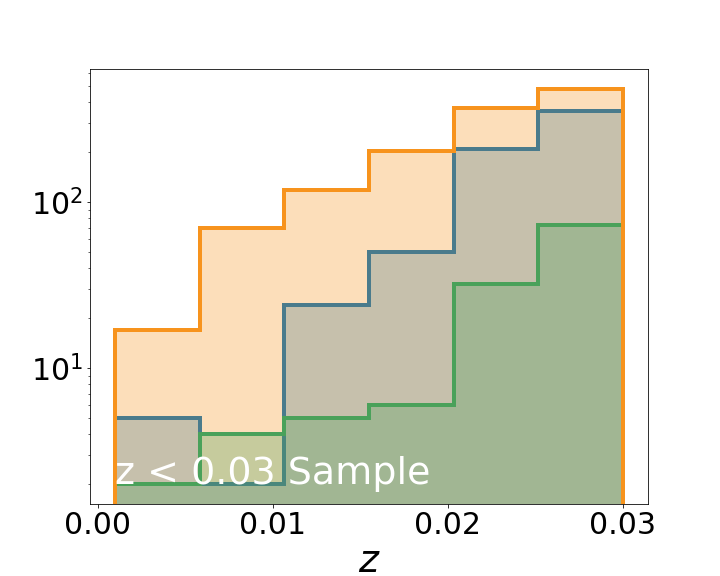}
\includegraphics[width=0.68\columnwidth]{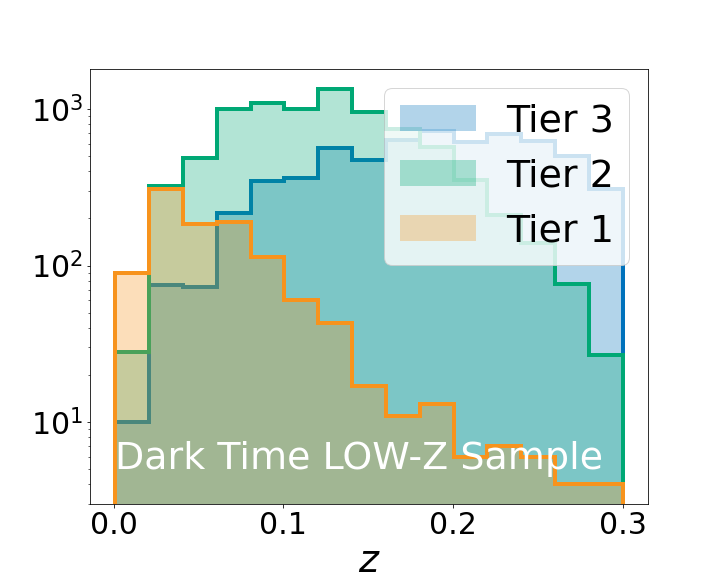}
\caption{\emph{Left}: Redshift distribution for all objects in the LOW-Z sample. The color represents which LOW-Z tier the objects come from. The redshift limits (0.001, 0.3) include $95\%$ of the full sample, with a small tail to higher redshifts. \emph{Center}: Redshift distribution for all objects between $0.001 < z < 0.03$. This redshift range is dominated by the Tier 1 objects selected by the CNN. \emph{Right}: Redshift distribution for all of the dark-time objects in the LOW-Z sample. The color represents which LOW-Z tier the objects come from. The redshift limits (0.001, 0.3) include $90\%$ of the full dark time sample, with a small tail to higher redshifts.} 

\label{fig:rshift_dist}
\end{figure*}

\begin{figure}[htb!]
\centering
\includegraphics[width=\columnwidth]{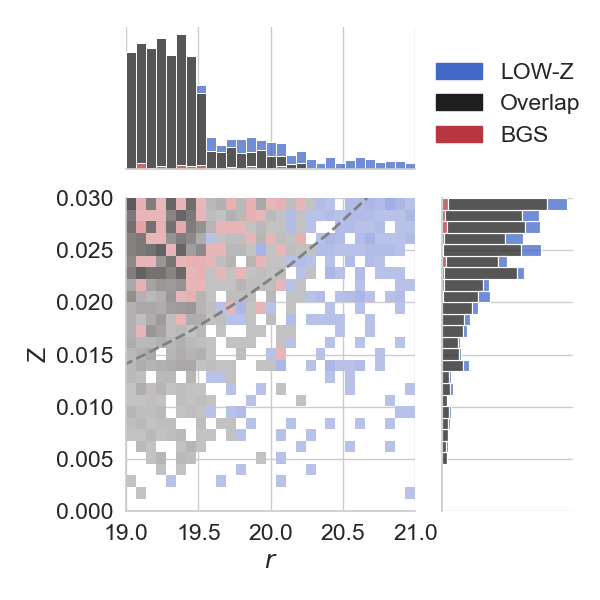}

\caption{Distribution of apparent $r$-band magnitude vs.\ redshift for the $z < 0.03$ galaxies from the LOW-Z (\emph{blue}) and BGS (\emph{red}) samples; objects belong to both samples (``overlap'') are plotted in \emph{dark grey}. The light grey dashed line marks the curve where $M_r = -15$. The histograms show the marginal distributions.} 

\label{fig:BGS_comp}
\end{figure}

\begin{figure*}[htb!]
\centering
\includegraphics[width=0.61\textwidth]{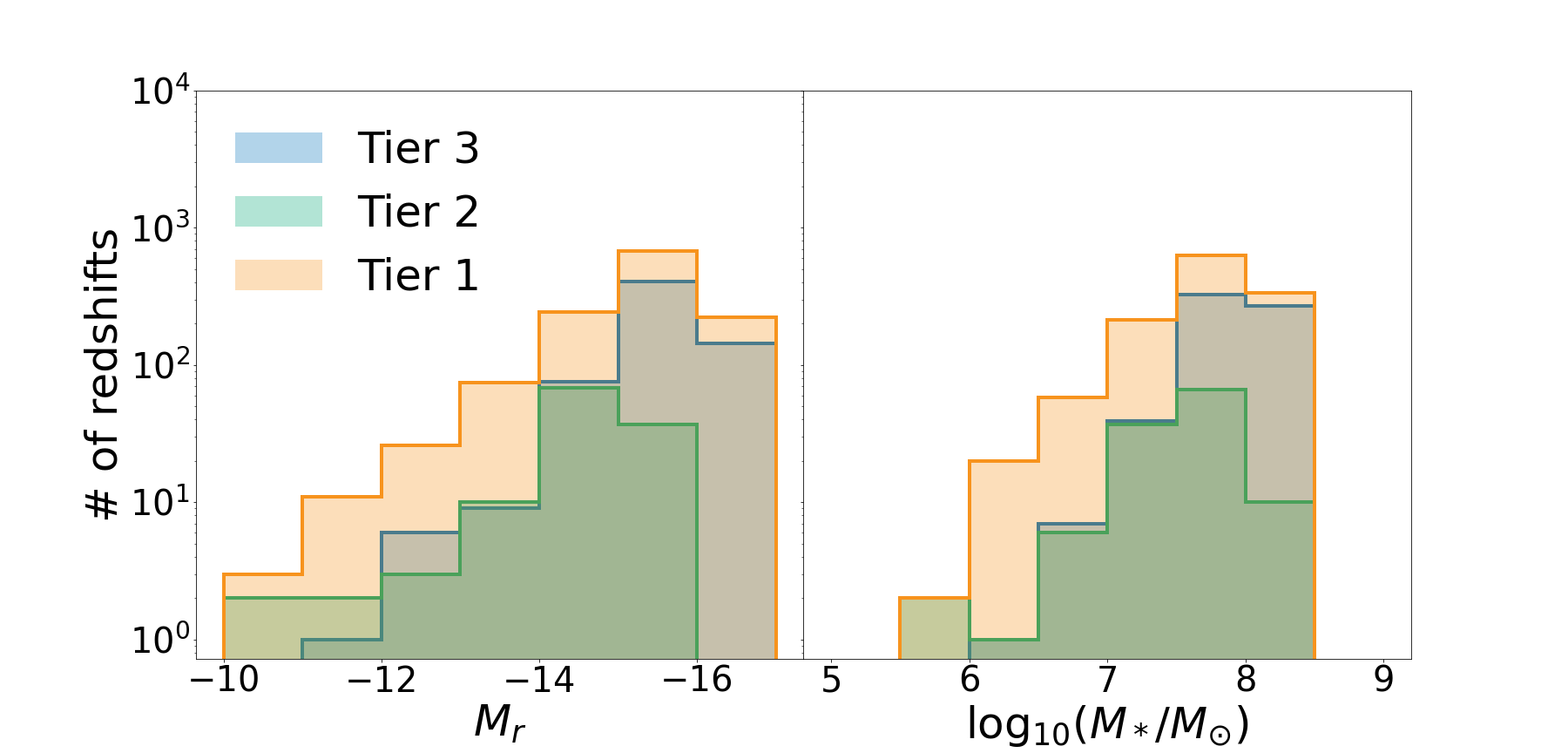}
\includegraphics[width=0.38\textwidth]{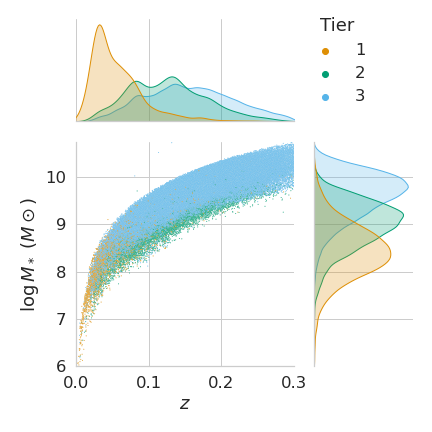}
\caption{\emph{Left}: Absolute $r$-band magnitude distribution for the LOW-Z sample at $z$ < 0.03 labeled by tier. \emph{Center}: Stellar mass distribution for the LOW-Z sample at $z$ < 0.03 labeled by tier. \emph{Right}: Scatter plot of redshift vs. $\log_{10}{M_*}$ $(\text{M}_\odot)$ colored by tier. The histograms on the top and left are normalized to show the shape of the distributions. By tier, the median redshift is [0.05, 0.12, 0.16] respectively, and the median stellar mass is [$10^{8.4}$, $10^{9.0}$, $10^{9.7}$] respectively.}

\label{fig:mag_dist}
\end{figure*}

\begin{figure*}[htb!]
\centering
\includegraphics[width=\textwidth]{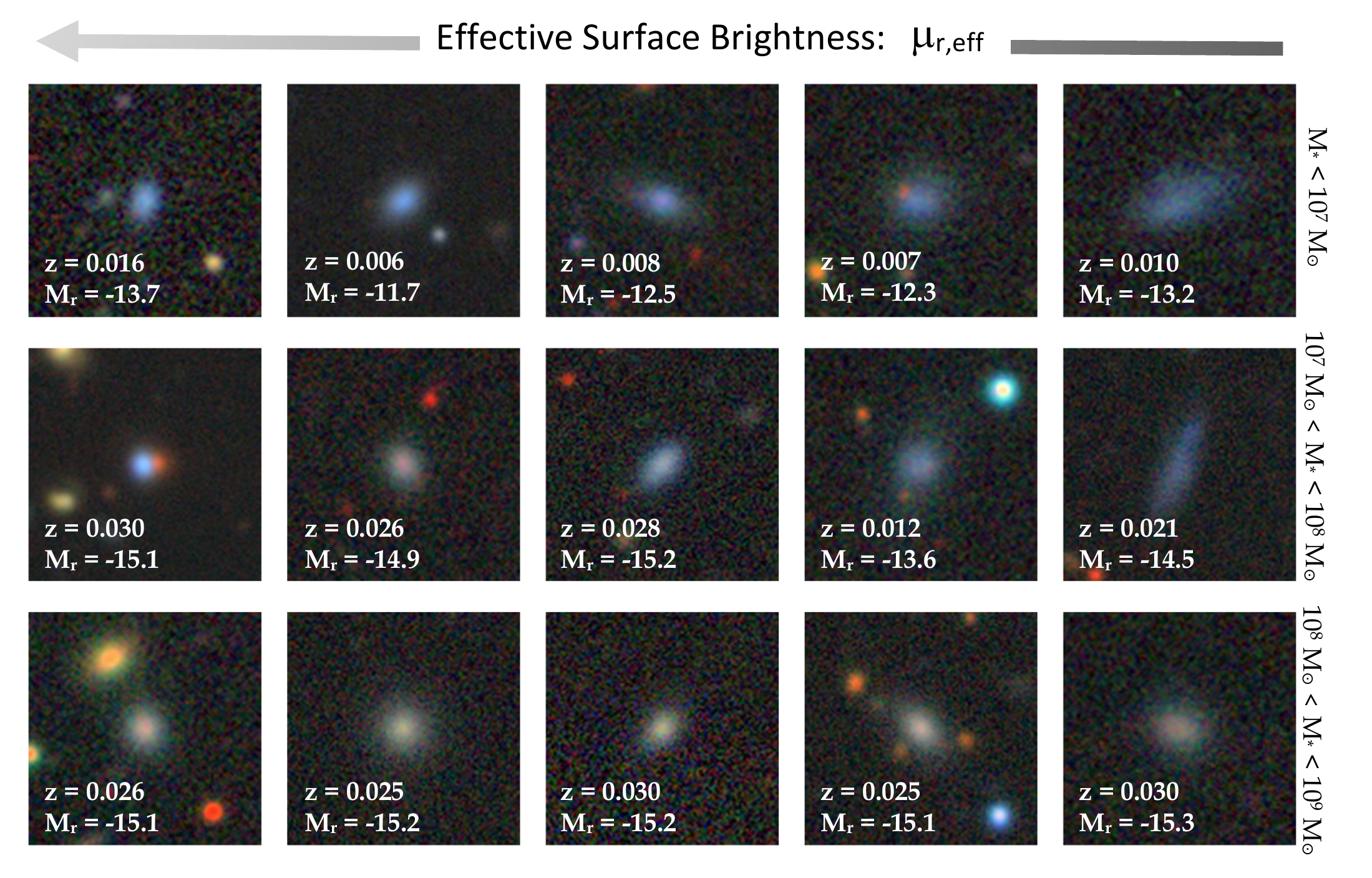}
\caption{Example observed LOW-Z objects with successful redshifts at $z<0.03$. Objects increase in stellar mass from top to bottom and decrease in surface brightness from left to right. The redshift and magnitude for each of the galaxies is labeled in white.} 

\label{fig:lowz_examples}
\end{figure*}

\begin{figure}[htb!]
\centering
\includegraphics[width=\columnwidth]{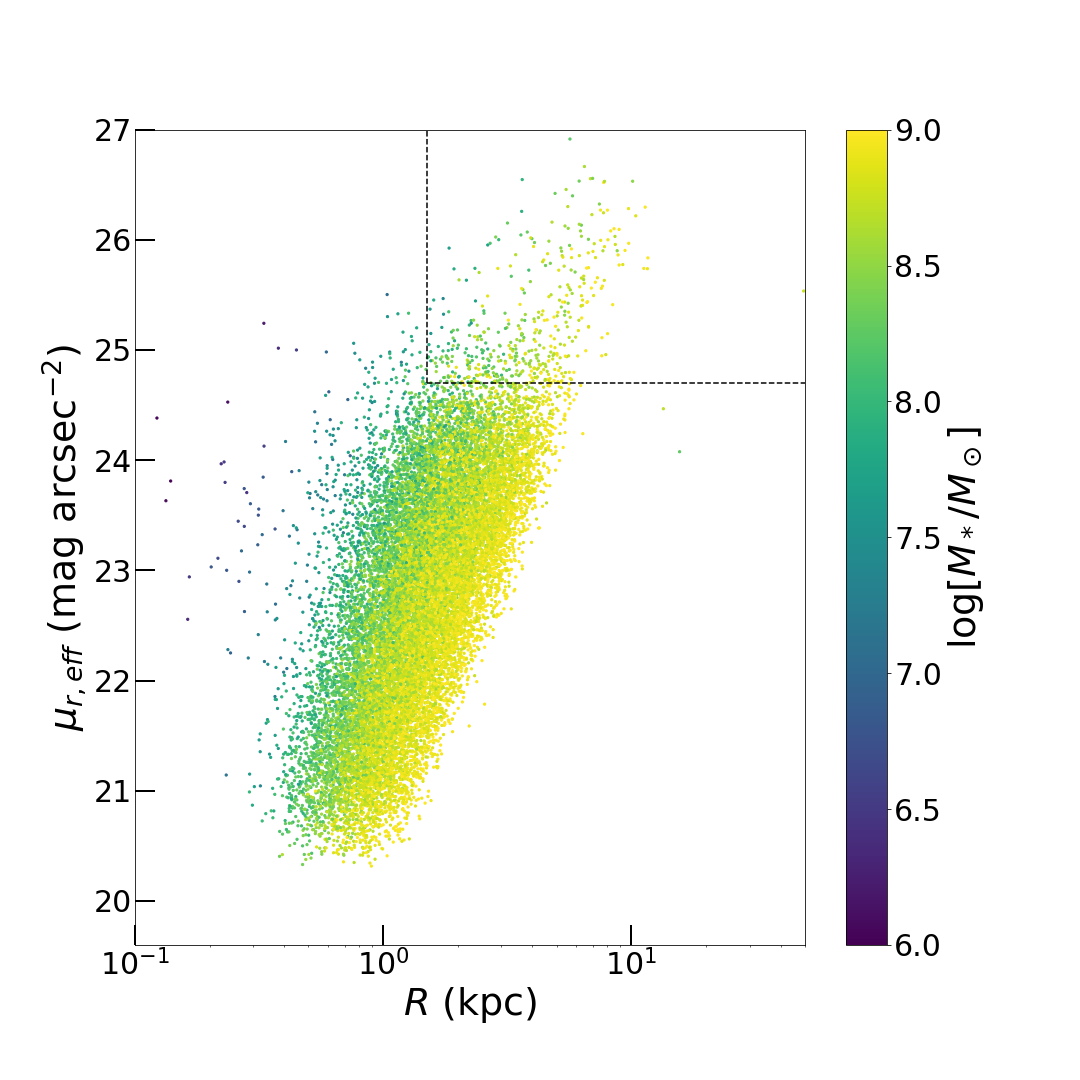}

\caption{Effective surface brightness vs.\ physical radius colored by stellar mass for galaxies with $M_* < 10^9$ M$_\odot$. Galaxies within the box in the upper corner are considered ultra-diffuse galaxies (UDGs).}

\label{fig:mu_r_r}
\end{figure}

\section{Characterizing the early LOW-Z Sample}
\label{sec:main_results}
The initial data for the LOW-Z program was taken between April -- June 2021 as part of the DESI One-Percent Survey (April -- May) and early Main Survey (May -- June). The early Main Survey data represents the first two months of data taken for year one (Y1) of the DESI survey and should be representative of the full Y1 dataset. The DESI One-Percent Survey took place before the beginning of data taking for Y1 of the DESI Main Survey. It was designed to operate similarly but with more passes per tile and longer exposure times. In total, the One-Percent Survey covered an area of 180 deg$^2$. The One-Percent dataset will be released as part of the Early DESI Data Release, expected in mid-2023 \citep{dr}. The LOW-Z targeting strategy was the same for both DESI One-Percent and Y1 (see Section \ref{sec:selection}). LOW-Z targets will be identifiable in all DESI Data Releases by selecting targets with: 
\begin{align*}
&\code{SCND\_TARGET} = 2^{15} \text{ (for tier 1)}; \\
&\code{SCND\_TARGET} = 2^{16} \text{ (for tier 2)}; \\
&\code{SCND\_TARGET} = 2^{17} \text{ (for tier 3)}. \\
\end{align*} 

\subsection{Redshift Sample By Tier}
Between the beginning of the DESI One-Percent Survey through the end of the scheduled summer shutdown in July, redshifts were obtained for 143,486 unique LOW-Z targets. Of the full sample, 6,633 are from Tier 1 ($4.6\%$), 9,479 are from Tier 2 ($6.6\%$), and the remaining 127,374 are from Tier 3 ($88.8\%$) (Table \ref{tab:dens_obs}). The Tier 3 objects dominate the sample due to the overlap with BGS. BGS targets have higher priority than LOW-Z and therefore have a higher fiber allocation fraction. Of the total number of objects that were allocated fibers in Tier 3, 120,002 ($94.2\%$) received fibers as part of the BGS sample (along with 5478 ($82.6\%)$ objects in Tier 1 and 724 (7.6\%) objects in Tier 2)\footnote{The small percentage of overlap with BGS in the Tier 2 sample is due to the use of an older definition of the BGS Faint color cuts at targeting.}. The redshift distribution of all three tiers can be seen in the left panel of Figure \ref{fig:rshift_dist}. This figure represents $96\%$ of the LOW-Z sample, with only $4\%$ of objects having redshifts $z > 0.3$ (the high-$z$ tail is not plotted for visual clarity). The median redshift of all objects in Tier 1 is $0.05$, for Tier 2, it is $0.12$, and for Tier 3, it is $0.15$. While this demonstrates the effectiveness of the whole LOW-Z program at selecting low-redshift galaxies, it especially exemplifies the efficacy of the CNN at selecting a sample of the lowest redshift objects ($z < 0.03$). 

Focusing on the lowest redshift objects ($z < 0.03$) in the LOW-Z sample, we are left with a sample of 2,019 objects: 1,257 are from Tier 1 ($62.3\%$), 122 are from Tier 2 ($6.0\%$), and 640 are from Tier 3 ($31.7\%$) (Table \ref{tab:dens_obs}). Approximately $20\%$ of the CNN-selected sample consists of objects at $z < 0.03$, consistent with the expected purity based on CNN cross-validation results. Of the $z < 0.03$ sample, 356 are faint ($r > 19.5$) non-BGS targets: 201 from Tier 1, 114 from Tier 2, and 41 from Tier 3. 

\subsection{LOW-Z Dark Time Sample and BGS Overlap}
We can also examine the sample of objects that received fibers specifically as part of the LOW-Z program (rather than BGS targets in the LOW-Z sample). These targets are interesting because they were observed in dark time, making it possible to get successful redshifts for fainter and lower surface brightness objects. Out of the 143,486 LOW-Z objects, 17,437 were observed during dark time. While 155 of these represent dark-time observations of BGS objects (due to overlap with the luminous red galaxies (LRG) or emission line galaxies (ELG)  samples), the rest are objects outside the BGS main sample (Table \ref{tab:BGS_cuts}). Out of the 17,437 objects, 1160 (6.6\%) are from Tier 1, 8757 (50.2\%) are from Tier 2, and 7520 (43.1\%) are from Tier 3. On average, the dark-time sample has slightly higher redshifts than the full sample. However, the median redshifts in Tier 1 and Tier 2 are the same as for the full sample, indicating this is mainly driven by the objects in Tier 3, which have a median redshift of 0.20. This result is likely due to the targeting error referenced in Section \ref{sec:summary}, meaning that the majority of Tier 3 objects in this regime are being sampled from the $z<0.03$-complete photometric cuts outside of both the $z<0.01$-complete color and surface brightness cuts. Since these objects are the reddest and most compact objects we target, we expect this sample to contain the lowest density of low-redshift objects. The full redshift distribution can be seen in the right panel of Figure \ref{fig:rshift_dist}. As shown in Figure \ref{fig:rshift_failure}, we are significantly more likely to obtain successful redshifts for low $r_\mathrm{fib}$ objects if they were observed in dark time. 

In Figure~\ref{fig:BGS_comp}, we also plot the apparent magnitude--redshift distribution for both the LOW-Z, BGS, and overlapping samples at $z < 0.03$ for the One-Percent and Early Main Survey data. About $80\%$ of the $z < 0.03$ objects are in the overlapping sample. A further $17\%$ of objects are exclusively LOW-Z galaxies; these galaxies tend to be fainter than the overlapping sample, as expected given the apparent magnitude range of the two surveys. The final $3\%$ of objects are exclusively BGS objects. These objects tend to be higher redshift and are discussed further in Section \ref{sec:selection_eff}.

\subsection{Galaxy Properties of the Early LOW-Z Sample}
The absolute $r$-band magnitude and stellar mass of the full LOW-Z sample at $z < 0.03$ is shown in the left and center panels of Figure \ref{fig:mag_dist}. K-corrected $r$-band absolute magnitudes are derived using the program \texttt{FastSpecFit}\footnote{\url{https://fastspecfit.readthedocs.io/en/latest/index.html}}. Stellar masses are derived using $g-r$ color and absolute $r$-band magnitude following \cite{mao2020saga}. While the distributions of Tier 2 and Tier 3 distributions look similar, the CNN-selected objects tend to be fainter in $M_r$ and at lower stellar masses. The LOW-Z sample contains a considerable number of galaxies with $M_* < 10^9$ $\text{M}_\odot$, making it an interesting data set for studying dwarf galaxies. Out of the full LOW-Z sample, 22,679 objects have $M_* < 10^9$ $\text{M}_\odot$, $2,011$ objects have $M_* < 10^8$ $\text{M}_\odot$, and 98 objects have $M_* < 10^7$ $\text{M}_\odot$. The right panel of Figure \ref{fig:mag_dist} shows the distribution of stellar mass as a function of redshift colored by tier. On average, both redshift and stellar mass increases as a function of tier, with Tier 1 objects making up the tail end of the stellar mass and redshift distribution (as can also be seen in the center panel of Figure \ref{fig:mag_dist}). The median stellar mass for Tier 1 is $10^{8.4}$ $\text{M}_\odot$ compared with $10^{9.0}$ $\text{M}_\odot$ and $10^{9.7}$ $\text{M}_\odot$ for Tiers 2 and 3 respectively. Example objects at $z<0.03$ can be seen in Figure \ref{fig:lowz_examples}, sorted by decreasing surface brightness and increasing stellar mass. Since stellar mass depends on color and surface brightness, higher stellar mass objects can be seen to be redder and have larger absolute magnitudes. 

The effective surface brightness vs.\ physical radius for the LOW-Z galaxies at $M_* < 10^9$ $\text{M}_\odot$ is shown in Figure \ref{fig:mu_r_r}. Of the LOW-Z dwarf galaxies, 469 are in the ultra-diffuse regime as defined by \cite{vandokkum2015}.

%\section{Understanding Effectiveness of LOW-Z Targeting Strategy} 
\subsection{LOW-Z Fiber Allocation}
\label{sec:one_percent}

As the LOW-Z program is a secondary target program, not all targets will be observed during the DESI survey. To understand the observed density of targets, we examine completed tiles taken as part of the One-Percent Survey. Out of the full sample, 36,481 objects received fibers during the One-Percent Survey, corresponding to 
%(2015 in Tier 1, 7445 in Tier 2, and 27,021 in Tier 3). This corresponds to 
an observed target density of $\sim$ 200 objects per square degree ([11, 41, 150] per deg$^2$ in Tier [1,2,3]; Table \ref{tab:tier_sum}) or approximately a $67\%$ fiber allocation fraction for the LOW-Z program. These numbers are consistent with close to $100\%$ fiber allocation for objects that overlap with the BGS survey and approximately $30\%$ fiber allocation for objects observed in dark time as part of the LOW-Z survey (Table \ref{tab:tier_sum}). Since we are a spare fiber program, the observed target density will vary over the sky depending on the density of the primary targets. However, because we are a dark time survey, our targets are primarily being displaced by ELG, LRG, and quasar targets. All of these surveys are focused on much higher redshift targets ($z > 0.4$), so our observed target density should not depend on the local density of objects at low-redshift but rather should vary approximately independently of the low-redshift environment across the sky. 

Since the One-Percent Survey had a different survey strategy than the main survey, which may have led to more LOW-Z targets receiving fibers than in the main survey, we verify these numbers using the DESI fiber assignment code \citep{fba} run on a small patch of the sky. After seven passes, we find that $\sim 30\%$ of our dark time targets are assigned fibers, while for BGS, after four passes in bright time we find that $\sim 75\%$ of targets are assigned fibers. This is lower than our estimate from the One-Percent Survey, which is most likely due to the extra passes per tile completed during the One-Percent Survey. Combining the fiber allocation across between bright and dark time gives a total fiber allocation for the LOW-Z survey of $\sim 50\%$.

Using data from the One-Percent Survey, we can also estimate the number of low-redshift ($z < 0.03$) targets per square degree we can expect to be observed as part of the LOW-Z Survey. During the One-Percent Survey, the LOW-Z sample selection returned 661 objects with $z < 0.03$. Since the One-Percent Survey covered $180^2$ degrees, this corresponds to approximately 3.7 observed objects per square degree (Table \ref{tab:tier_sum}). This number may be a slight overestimation for the full survey as the One-Percent Survey had longer exposures and more passes per tile than the main survey.
%, meaning our targets were more likely to be allocated fibers despite their low priority. 

\begin{figure*}[htb!]
\centering
\includegraphics[width=\textwidth]{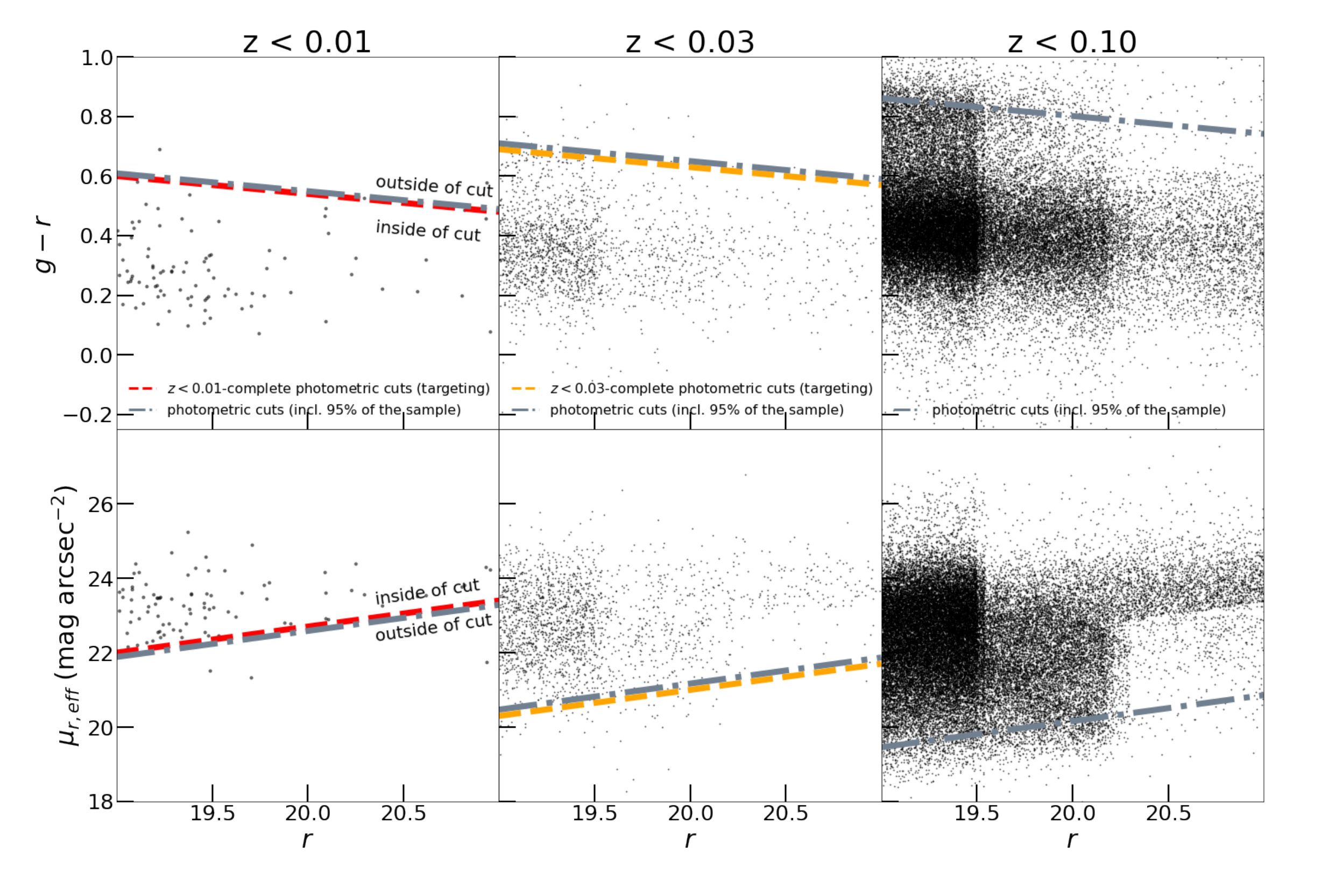}
\caption{Threshold as a function of redshift constrained to include 95\% of the galaxies in the DESI data. The slope is the same as for the $z<0.01$-complete and $z<0.03$-complete photometric cuts (Equations \ref{eq:targeting-cuts-gr-r}; \ref{eq:targeting-cuts-sb-r}). The grey dash--dotted line marks the threshold that incorporates 95\% of the galaxy sample for the distribution of $g - r$ color (\emph{top}) and surface brightness (\emph{bottom}) for the combined sample (shown in black). For reference, the $z<0.01$-complete photometric cuts (\emph{red dashed}) and $z<0.03$-complete photometric cuts (\emph{orange dashed}) are plotted as well. The objects outside of the LOW-Z $z<0.03$-complete photometric cuts are dominated by the DESI Bright Galaxy Sample.}
\label{fig:target_fit}
\end{figure*}

\begin{figure*}[htb!]
\centering
\includegraphics[width=\textwidth]{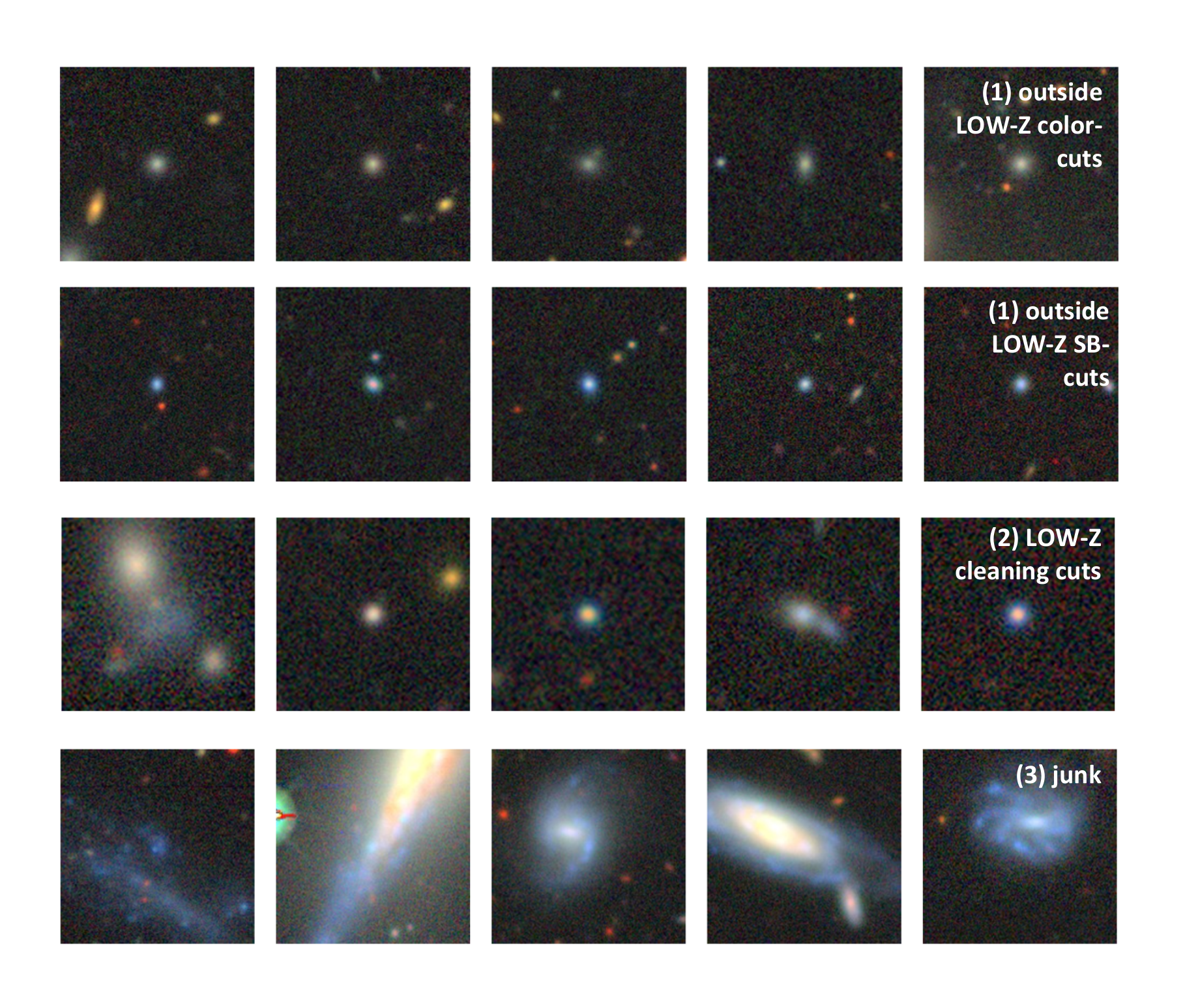}
\caption{Example DESI objects at $z$ < 0.03 that are not in the LOW-Z sample. \emph{(1) Top:} Objects that are outside of the $z<0.03$-complete color cuts. \emph{(2) Second Row:} Objects that are outside of the $z<0.03$-complete surface brightness cuts. 
\emph{(3) Third Row:} Objects that are within the $z<0.03$-complete photometric cuts but are removed by our photometric cleaning cuts. \emph{(4) Bottom:} junk objects (generally parts of larger galaxies).}
\label{fig:missed_objects}
\end{figure*}

\begin{figure*}[htb!]
\centering
\includegraphics[width=.9\columnwidth]{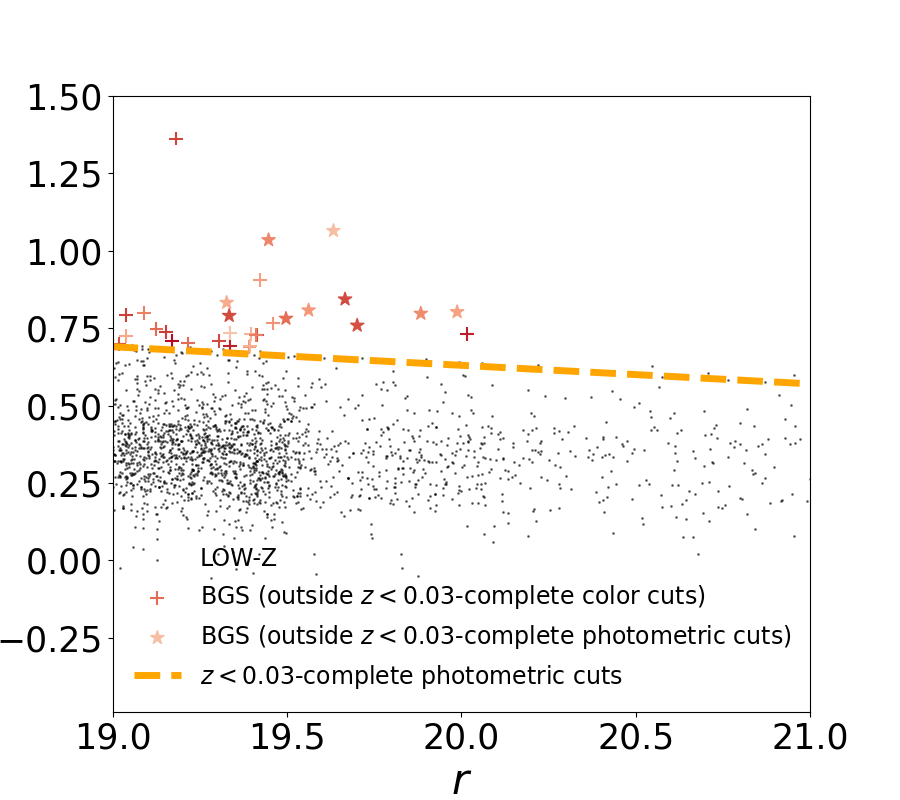}
\includegraphics[width=1.1\columnwidth]{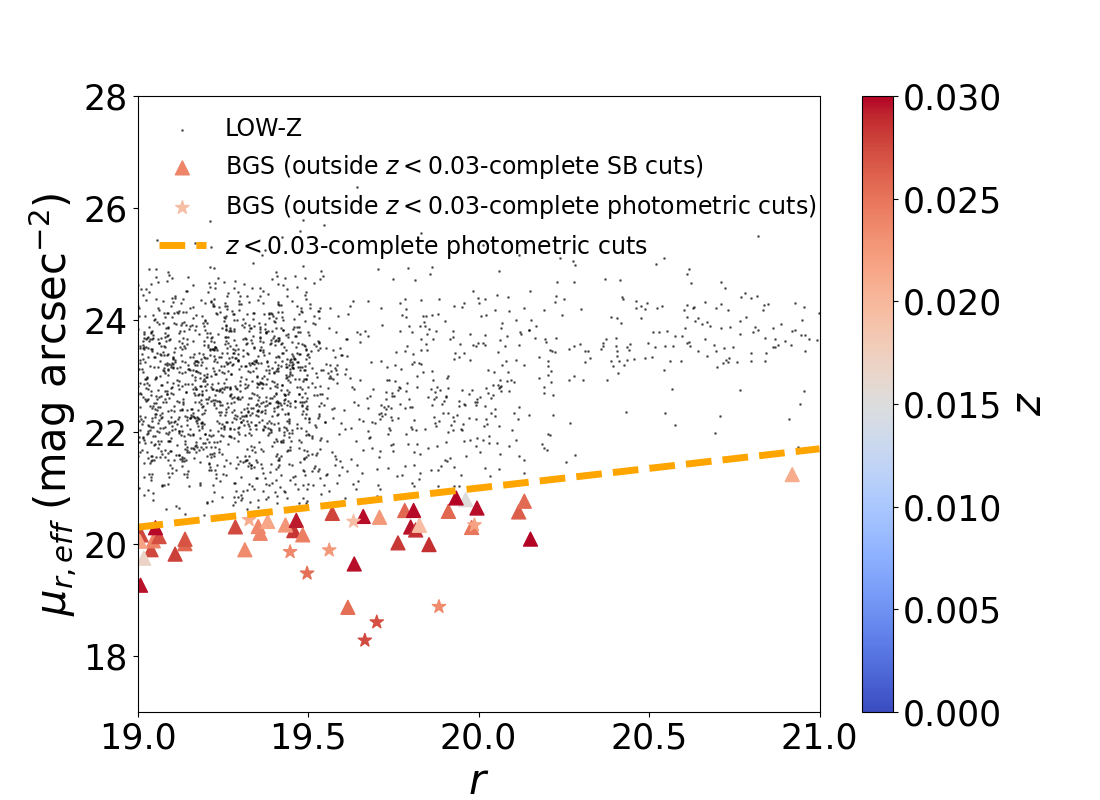}
\caption{Magnitude--color (\emph{left}) and magnitude--surface brightness (\emph{right}) plots for the $z$ < 0.03 objects outside of the $z<0.03$-complete photometric cuts colored by redshift. All of these objects come from the BGS sample. The color--surface brightness distributions of the LOW-Z objects within the $z<0.03$-complete photometric cuts at $z$ < 0.03 are plotted in black. }
\label{fig:missed_g_r_mu}
\end{figure*}

\subsection{Sample Selection Validation}
\label{sec:selection_eff}

We further validate our sample selection methods by using all redshifts from DESI Y1 data, including redshifts that are not from the LOW-Z program. As these objects are not part of the LOW-Z photometric sample, it allows us to examine if any low-redshift objects are missed from our sample selection. Due to the overall design of the DESI survey, at low redshifts ($z < 0.1$), this sample is dominated by galaxies from the BGS galaxy sample. While the BGS sample is only complete out to $r = 19.5$, it is not subject to the same color and surface brightness cuts as the LOW-Z sample, allowing us to validate the completeness of the current set of catalog-level photometric cuts for a sample of objects outside of the LOW-Z selection. However, this calculation is limited by the dominance of the LOW-Z sample at low redshifts in the DESI data as well as underlying incompleteness in the Legacy Imaging DR9 photometric catalogs used to select all DESI targets. We discuss these limitations further at the end of this section. 

Figure \ref{fig:target_fit} shows the low-redshift galaxies in DESI as a function of $r$, $g - r$, $\mu_{r, {\rm eff}}$, and redshift. The grey line is constrained to have the same slope as the $z<0.01$-complete and $z<0.03$-complete photometric cuts, and the intercept is varied to capture $95\%$ of the sample. We see a steady redshift-dependent evolution for both the $g - r$ and $\mu_{r, {\rm eff}}$ fits. We recover that the $z<0.01$-complete photometric cuts are complete to the SAGA goal of $z < 0.01$. Furthermore, the difference between the fit 
%(\emph{grey dashed--dotted line}) 
and the $z<0.03$-complete photometric cuts 
%(\emph{orange dashed line}) 
is negligible for both $g - r$ and surface brightness. This indicates that the $z<0.03$-complete photometric cuts are indeed quite complete out to the LOW-Z sample goal of $z < 0.03$ relative to the broader cuts used to select BGS galaxies.

Focusing on the $z < 0.03$ objects that are not in the LOW-Z sample, we can divide the objects into three categories: (1) objects that are outside the LOW-Z color--surface brightness cuts, (2) objects that are within the LOW-Z color--surface brightness cuts but that were excluded from the LOW-Z sample due to our catalog cleaning cuts (Section \ref{sec:cleaning_cuts}), and (3) junk objects. The third category mainly consists of misclassified pieces of brighter galaxies. Examples of objects from each of the three categories can be seen in Figure \ref{fig:missed_objects}. 

The objects in the first and second categories tend to be compact, high-surface brightness objects, while the objects in category three tend to be miscentered large, bright nearby galaxies. Out of the 994 objects with DESI spectra at $z$ < 0.03 that are not part of the LOW-Z survey, only 97 are outside of the $z<0.03$-complete photometric cuts. This aligns with what we see in Figure \ref{fig:target_fit} that $\sim 95\%$ of the sample at $z < 0.03$ is within the $z<0.03$-complete photometric cuts. These 97 objects represent less than $5\%$ of the sample at $z < 0.03$. A further 397 are within the $z<0.03$-complete photometric cuts but are removed from the sample due to the photometric cleaning cuts imposed in Section \ref{sec:cleaning_cuts}. The majority of these objects are removed by the cut on \code{SIGMA\_GOOD} $\geq 30$ and \code{RCHISQ} $\leq 0.85$. As a result, we modify our target selection in Y2 so that they do not include these requirements (see Section \ref{sec:desiii} for details). The remaining are junk objects. 

In order to better understand the objects that were being missed outside of the $z<0.03$-complete photometric cuts, we visually inspected the spectra of the 97 objects. Twenty of these objects were found to be either quasars or stars misidentified as galaxies. Another six were objects at $z \sim 0.1$ misclassified as $z < 0.03$. Removing these objects left us with 71 objects that were actually galaxies at $z < 0.03$ outside of the $z<0.03$-complete photometric cuts, all of which come from the BGS sample. Out of the 71 objects, 20 are outside of the $z<0.03$-complete color cuts, 41 are outside of the $z<0.03$-complete surface brightness cuts and a further ten are outside of both the color and surface brightness cuts. The color--surface brightness distribution of objects can be seen in Figure \ref{fig:missed_g_r_mu}. The vast majority of these objects are at $z > 0.02$ (66 out of 71). 

We further investigate the 30 objects that fall outside of our $z<0.03$-complete color cuts. These objects are of particular interest as we want a complete sample of quenched objects in order to further understand dwarf galaxy formation as a function of environment with the LOW-Z sample. Of these objects, 28/30 are at $z > 0.02$, and all of them are at $z > 0.01$. Four of these objects are blended objects with incorrect photometry. Most of the remaining are in extremely high-density environments (13/30 are members of the Coma Cluster), where we expect to find the reddest and most compact low-mass objects.

Removing all objects in Coma and with obvious photometric errors, we are left with only thirteen objects. While these objects represent an interesting sample for further follow-up, they do not indicate a significant population of isolated quenched objects outside of our $z<0.03$-complete color cuts. 

As stated above, this analysis is limited by the sample of redshifts available in DESI, which is dominated by objects selected by the LOW-Z program. Since LOW-Z is pushing the forefront for faint low-redshift surveys, accurate characterization of its completeness is difficult given the lack of available data with which to compare; externally validating our redshift completeness will remain an active area of research for the program going forward. We are also limited by catalog-level incompleteness in the Legacy Imaging Survey DR9 catalogs used to select DESI targets, especially for the lowest surface brightness objects. We anticipate that these biases will be better characterized by current \citep[e.g.,][]{HSC-SSP,Danieli_2020,Carlsten_2022} and future low surface brightness galaxy surveys \citep[e.g.,][]{Roman,LSST,Euclid_LSB}, and partially ameliorated by more advanced techniques for constructing and cleaning photometric catalogs \citep[e.g.,][]{Walmsley_2019,Greco_2021,Tanoglidis_2021,DiTeodoro_2023}.

In addition to incompleteness in our target selection, an additional source of incompleteness comes from observed sources for which we are unable to accurately determine a redshift. We discuss redshift failure rates further in the following section. However, since we do not see evidence for a population of galaxies we are missing with our current $z<0.03$-complete photometric cuts, we do not propose an update to the photometric selection for DESI Year 2 (Section \ref{sec:Y2}).

\begin{figure*}[htb!]
\centering
\includegraphics[width=\textwidth]{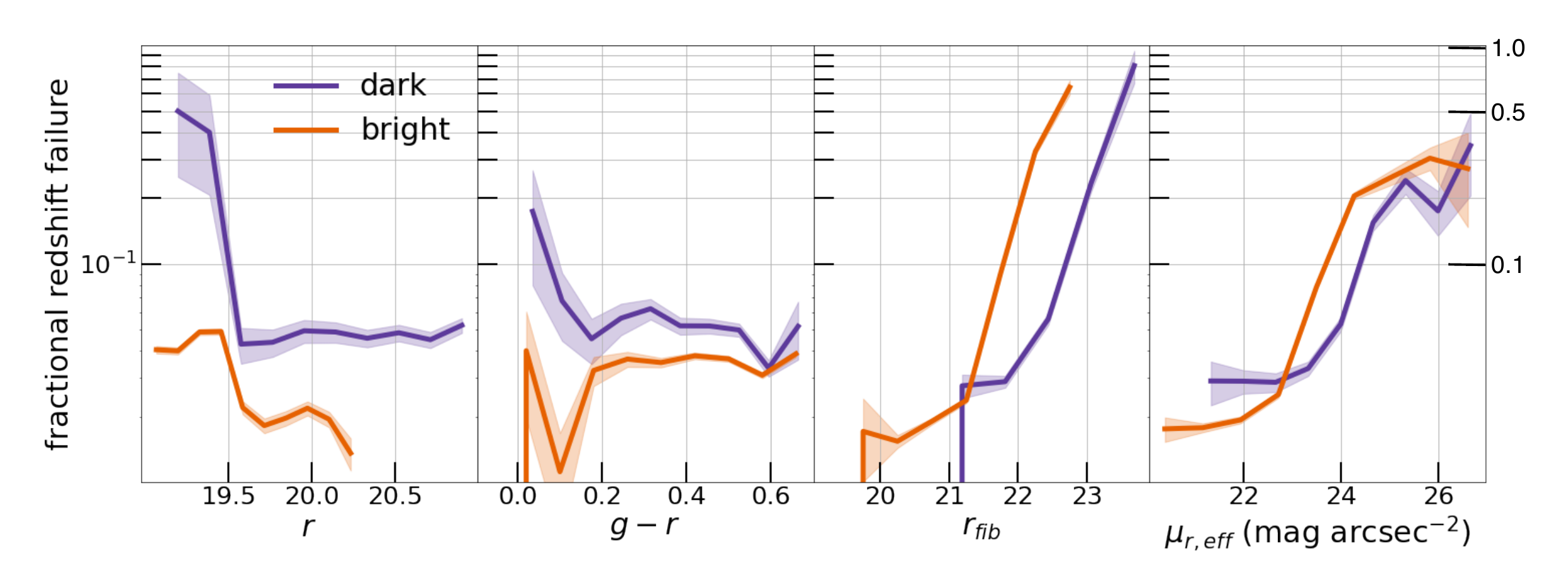}
\caption{Fractional redshift failure as a function of $r$ (\emph{left}), $g - r$ (\emph{center left}), $r_\mathrm{fib}$ (\emph{center right}), and $\mu_{r, {\rm eff}}$ (\emph{right}). The sample is split between spectra obtained in bright time (\emph{red}) and dark time (\emph{purple}) with Poisson error bands. The increase in redshift failures in dark time at r < 19.5 is being driven by the low fiber magnitude of these large nearby objects, all of which are beyond the BGS fiber magnitude cut at $r_\mathrm{fib} = 22.9$.} 
\label{fig:rshift_failure}
\end{figure*}

\begin{figure}[htb!]
\centering
\includegraphics[width=\columnwidth]{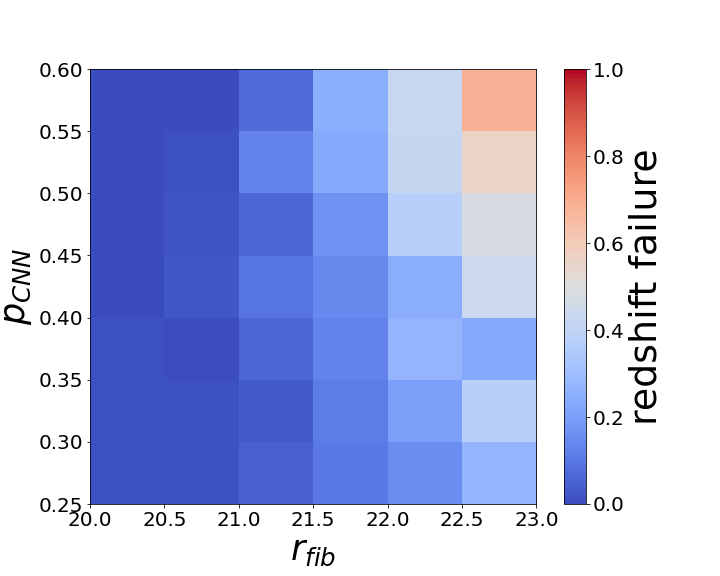}
\caption{Fractional redshift failure as a function of $r_\mathrm{fib}$ and $p_{\rm CNN}$. Objects with $p_{\rm CNN} > 0.6$ are assigned a value of $p_{\rm CNN} = 0.6$ to increase statistics in the highest bin.} 
\label{fig:rshift_failure_pCNN}
\end{figure}

\begin{figure*}[htb!]
\centering
\includegraphics[width=\columnwidth]{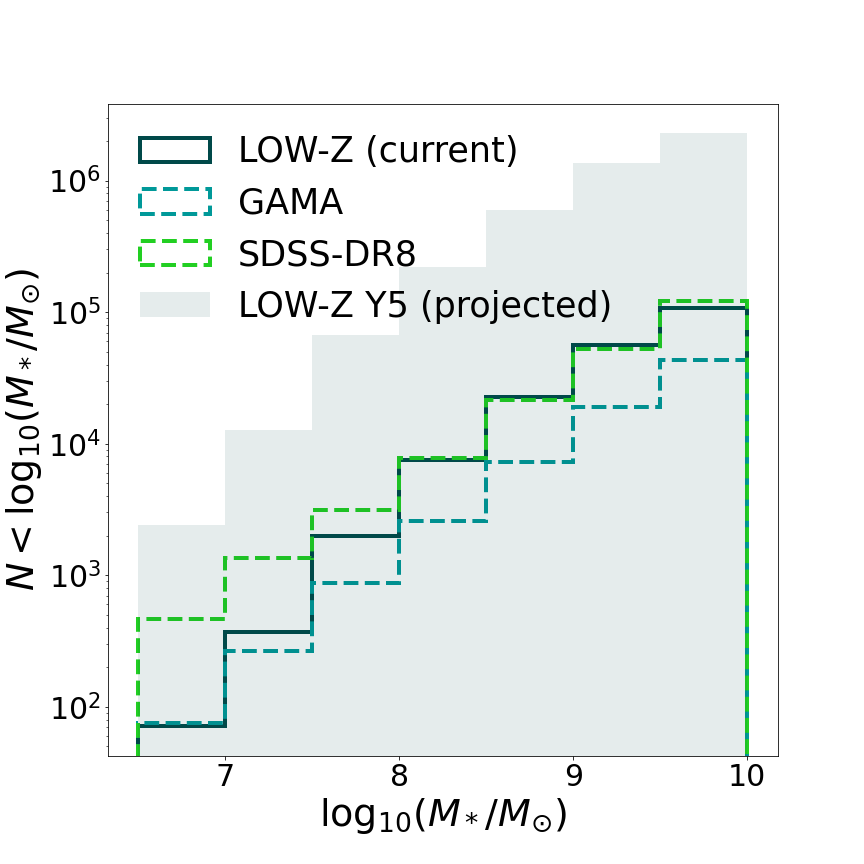}
\includegraphics[width=\columnwidth]{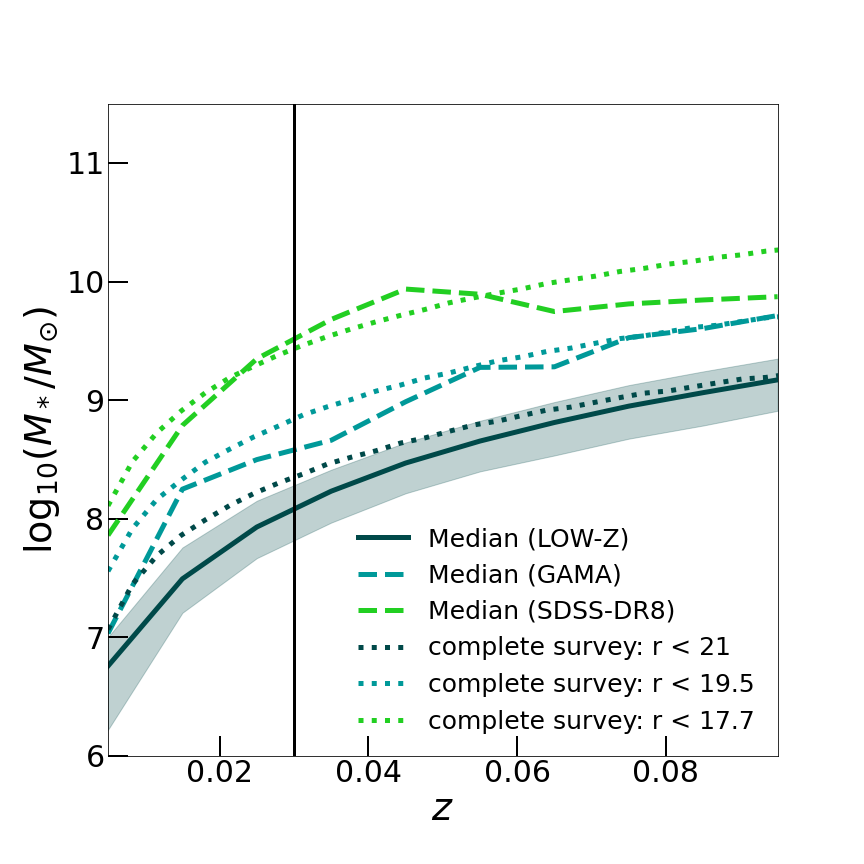}
\caption{\emph{Left}: Comparison of cumulative number of galaxies at $N < M_*$ between $10^{6.5}$  M$_\odot < M_* < 10^{11}$  M$_\odot$ for LOW-Z (\emph{dark blue}), GAMA (\emph{light blue}), and SDSS-DR8 (\emph{green}). The dark blue shaded region gives an estimate for the LOW-Z survey after the full five-year run by rescaling the One-Percent Survey data to the full $14,000$ deg$^2$ footprint. GAMA covers 250 deg$^2$ and is complete to $r < 19.65$. SDSS-DR8 covers 9380 deg$^2$ and is complete to $r < 17.77$. Objects are restricted to $M_* > 10^{6.5}$  M$_\odot$ due to irregularities in the stellar mass catalog from SDSS-DR8. \emph{Right}: median stellar mass as a function of redshift at $z < 0.1$ for the LOW-Z sample (\emph{dark blue}), the GAMA sample (\emph{light blue}), and the SDSS-DR8 sample (\emph{green}). For visual comparison, the dotted lines represent the median stellar mass for a complete survey down to $r < 21$ (\emph{dark blue}),  $r < 19.5$ (\emph{light blue}), and $r < 17.7$ (\emph{green}) roughly corresponding to the magnitude limits for LOW-Z, GAMA (or BGS), and SDSS-DR8 respectively. The shaded region marks the 68\% sample variance in $M_*$ as a function of $z$ for the LOW-Z sample; $z = 0.03$ is marked by the black line.}

\label{fig:mstar_comp}
\end{figure*}

\subsection{Redshift Success Rate}
\label{sec:rshift_failure}
Since our sample extends to fainter $r$-band apparent magnitudes than the BGS sample, we are interested in the redshift success rates for these objects. This has important implications both for understanding the power of the DESI instrument as well as understanding the completeness of the observed LOW-Z sample. We define a successful redshift as a redshift as $\texttt{ZWARN} = 0$ and $\Delta \chi^2 > 30$, indicating no warning flags raised and a high level of redshift confidence ($\Delta \chi^2$ is the difference in $\chi^2$ for the two best-fitting models). Redshift failure as a function of $r$, $g - r$, $\mu_{r, {\rm eff}}$, and $r_\mathrm{fib}$ is shown in Figure \ref{fig:rshift_failure}. We separate out bright and dark time observations to show the dependence of the failure rate on observing conditions. However, we do not separate between observations taken in Y1 and the One-Percent Survey. Despite the differences in survey strategy and longer exposure times for the One-Percent data, we do not find a significant difference in the redshift failure rates as a function of any of the variables we consider between DESI One-Percent and Y1, leading to our choice to show results from the combined sample. 

Redshift failure rates show negligible evolution as a function of $r$ and $g - r$ for both bright and dark time targets, indicating that DESI is able to capture redshifts out to our apparent magnitude limit of $r = 21$ in dark time and out to the BGS limit of $r = 20.2$ in bright time across the full range of $g - r$ colors included in the $z<0.03$-complete color cuts. The increase in redshift failures in dark time at $r < 19.5$ is due to the fact that the only objects in this regime observed in dark time are objects outside of the BGS $r_\mathrm{fib}$ cut and thus correspond to a sample of objects with $r_\mathrm{fib} > 22.9$. Therefore, the increasing failure rate can be attributed to their high $r_{\rm fib}$ rather than a magnitude dependence. We do see a significant increase in redshift failures for the lowest surface brightness and $r_\mathrm{fib}$ objects. The failure rate increases to almost $40\%$ at $\mu_r = 27$ mag arcsec$^{-2}$ for both dark- and bright-time targets. The trend with $r_{\rm fib}$ is even more dramatic, with failures increasing to around $70\%$ at the bright-time limit of $r_\mathrm{fib} = 22.9$ for objects observed in bright time and to a similar rate at $r_\mathrm{fib} = 24$ for objects observed in dark time. This indicates that object surface brightness and, by extension, fiber magnitude rather than apparent magnitude is the biggest limitation for getting successful LOW-Z redshifts with DESI. In addition, it can be seen in Figure \ref{fig:rshift_failure_pCNN} that redshift failure is correlated with $p_{\rm CNN}$ at fixed $r_\mathrm{fib}$, which is statistically driven by bright time observations (the dark time spectroscopic failure rates do not show a significant trend). This is concerning because it indicates that redshift failure rates may be correlated with the likelihood of an object being low redshift. Observational effects may be able to explain this trend: the spectra of lower-redshift galaxies feature the [O\,\textsc{ii}] doublet emission line at bluer observed wavelengths, where the DESI spectrograph is less sensitive \citep[see, e.g.,][]{DESI2022}.\footnote{Based on Figure~27 of \citet{DESI2022}, we estimate that the filter transmission for [O\,\textsc{ii}] observed wavelength decreases by roughly a factor of two between redshift $z=0.15$ to $z=0$.} Thus, it becomes more difficult to confirm the redshift for \textit{bona fide} lower-redshift galaxies via the distinguishing [O\,\textsc{ii}] spectral feature. If $p_{\rm CNN}$ truly selects lower-redshift galaxies, then we may expect targets with higher values of $p_{\rm CNN}$ to result in a higher rate of redshift failures. 
We expect to be able to characterize this potential effect significantly better using the full Y1 and Y2 datasets.

\begin{figure}[t]
\centering
\includegraphics[width=\columnwidth]{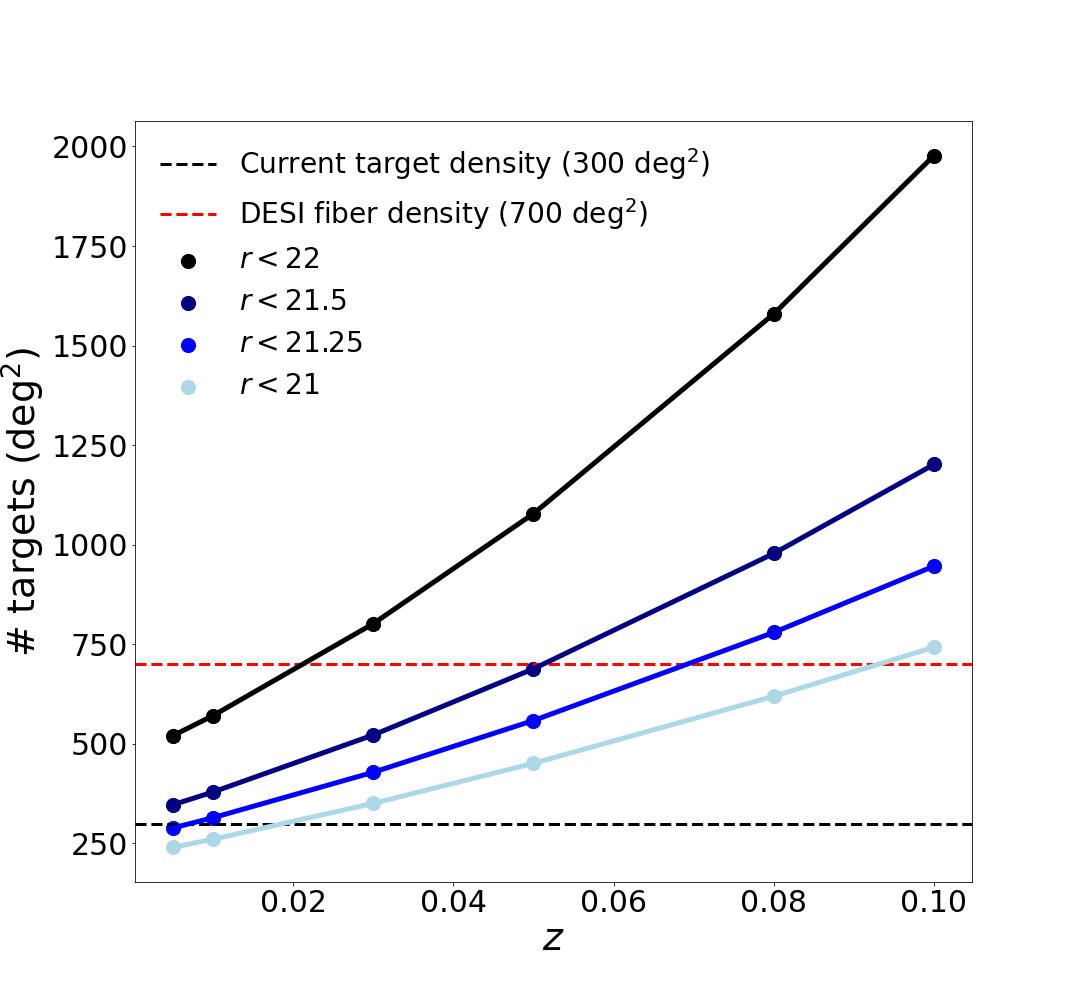}
\caption{Predicted target density as a function of redshift for different $r-$band apparent magnitude cuts: 21 (\emph{cyan}), 21.25 (\emph{light blue}), 21.5 (\emph{blue}), and 22 (\emph{navy}). The black dashed line indicates the current LOW-Z target density, while the red dashed line indicates the approximate target density of a single pointing of the DESI spectrograph.} 
\label{fig:density}
\end{figure}

\section{Discussion: The LOW-Z Survey in Context} 
%\subsection{The LOW-Z Survey in Context}
The sample of low-redshift galaxies from LOW-Z is already significant when compared to previous surveys. The SDSS main survey (covering $9380$ deg$^2$) only has $\sim 0.5$ objects per square degree at $z < 0.03$ and only a few hundred are at $r > 19$ \citep{Aihara2011}. GAMA, meanwhile, has about 17 objects per square degree at $z < 0.03$ but only covers $250$ square degrees of the sky \citep{Driver2022}. With more than 2000 objects at $z<0.03$, the LOW-Z sample is already competitive with the GAMA and SDSS samples, which each contain around 5000 objects at $z<0.03$. Additionally, SDSS is only complete down to $r = 17.77$ and GAMA to $r = 19.65$. The LOW-Z sample also already contains roughly the same number of objects at $r > 19$ as the SAGA sample. SAGA has 1,440 objects at $z < 0.03$ (420 of which are at $r > 19$). This corresponds to approximately 17 (5 at $r > 19$) objects per square degree (because the number density is enhanced by satellite galaxies around SAGA hosts).

We use the GAMA luminosity function \citep{Loveday150501003} to estimate the total expected density of $z < 0.03$ objects in the sky to a given magnitude. At $r < 21$, we expect approximately 16 objects per square degree (8 per deg$^2$ between $19 < r < 21$). For the LOW-Z sample, we find an observed density within this magnitude and redshift range of 3.7 objects per square degree (Table \ref{tab:tier_sum}). Correcting for the low fiber allocation fraction of the LOW-Z program (approximately 67\% in the One Percent Survey; Section \ref{sec:one_percent}), this gives us an estimated completeness of $\sim 70\%$. This is slightly lower than our estimate of $ > 95\%$ target completeness for the $z<0.03$-complete photometric cuts (Section \ref{sec:selection_eff}). This underestimate could be due to the over-representation of BGS objects in the One-Percent sample (see Section \ref{sec:one_percent}), which skews the sample to brighter magnitudes where we expect a lower number density of $z<0.03$ objects, due to uncertainties about our completeness (see Section \ref{sec:selection_eff}), or due to sample variance in the GAMA estimate. For comparison, at $z < 0.03$, BGS Bright has about 1.5 objects per square degree at $r < 19.5$, and BGS Faint has about $0.5$ between $19.5 < r  < 20.3$. Assuming $> 95\%$ fiber allocation for BGS during the One Percent Survey, we estimate that the BGS target selection is close to $100\%$ complete at $r < 19.5$ and $15\%$ complete between $19.5 < r < 20.3$ for objects at $z < 0.03$. Figure \ref{fig:BGS_comp} already shows how the LOW-Z survey complements the BGS Faint sample by filling in the lowest redshift galaxies between $19.5 < r < 20.3$. 

A comparison between the number of galaxies at $M < M_*$ between LOW-Z, GAMA \citep{Driver2022}, and SDSS--DR8 \citep{Aihara2011, Kauffmann2003, Blanton2011} is shown in the left panel of Figure \ref{fig:mstar_comp}. At $M_* < 10^9$ M$_\odot$, LOW-Z already has more galaxies than GAMA and is competitive with SDSS. The grey-shaded region gives the estimated number of LOW-Z galaxies that would be observed over the full 14,000 deg$^2$ DESI footprint. This estimate is done by rescaling the number of galaxies in the 180 deg$^2$ region covered by the One-Percent Survey to the full survey area and does not account for the updates in targeting described in Section \ref{sec:Y2}. Even in the lower limit where no targeting improvements are included, we predict that by the end of the five-year DESI survey, if the LOW-Z survey continued as it did in Year 1, it will have surpassed the number of dwarf galaxies ($M_* < 10^9$ M$_\odot$) identified by the SDSS main survey and GAMA by an order of magnitude.  

The right panel of Figure \ref{fig:mstar_comp} shows the median stellar mass as a function of redshift for the LOW-Z, GAMA, and SDSS-DR8 samples. For comparison, the dashed lines show the median redshift for a complete magnitude-limited survey assuming the GAMA luminosity function and a luminosity--stellar mass relation fit to the GAMA data. The LOW-Z sample has a lower median stellar mass at all redshifts than either the GAMA or SDSS-DR8 samples and lies close to the theoretical line for a complete magnitude-limited survey to $r < 21$.

%With DESI-II planning already getting started, the LOW-Z survey is an optimal testing ground for exploring how to design an optimally targeted low redshift survey. Using Equations \ref{eq:g-r_fit} and \ref{eq:mu-r_fit} and photometric data from DR9, we can estimate target density of a theoretical complete LOW-Z survey to a given target redshift and apparent magnitude.

\section{The Future of LOW-Z}
\label{sec:desiii}

The LOW-Z program will continue to survey a highly complete sample of $z < 0.03$ objects. 
Using our results from the survey validation and the first two months of DESI Y1 observations, we implement the following updates to the LOW-Z survey targeting strategy for DESI Y2. These updates (a) reduce overlap with BGS, and (b) improve the completeness for the more efficient CNN selection. Based on Figure \ref{fig:rshift_failure}, we extend the faint end of the LOW-Z survey to $r = 21.15$. However, in combination with this extension, we implement a fiber magnitude cut at $r_{\rm fib} < 23.5$ to avoid targeting objects with a low likelihood of redshift success. We also remove the cleaning cut mentioned in Section \ref{sec:selection_eff}. Additionally, we remove objects that overlap with the BGS survey given the high fiber allocation fraction and redshift success BGS has achieved so far \citep{Hahn2022}. Since we are removing the BGS targets, we expect the Y2 LOW-Z fiber allocation fraction to be lower than that found for Y1; we estimate that it will be $\sim 30\%$, with all of these objects receiving fibers in dark time. The exception will be objects with $r_\mathrm{fib} > 22$, where we expect to achieve higher redshift success in dark time (Figure \ref{fig:rshift_failure}). We expect that this change, which allows us to include all objects in the $z<0.03$-complete photometric cuts without subsampling, to maximize the number of $z < 0.03$ objects targeted by DESI between the BGS and LOW-Z samples. Combined, these changes only require a slight increase to the LOW-Z target density ($\sim 425$ targets per deg$^2$).\footnote{These updates to the target selection occurred during the planned summer shutdown between Y1 and Y2. The remainder of the Y1 sample --- not analyzed here --- was targeted according to the tiers laid out in Section \ref{sec:summary}.}

\begin{deluxetable*}{l c c c}
\centering
\tablecaption{\label{tab:cnn_y2} CNN forecasts for Y2 LOW-Z Tier 1}
\tablehead{Sample & Target density [deg$^{-2}$] & \colhead{Estimated completeness ($z < 0.03$)} & \colhead{Number of $z < 0.03$ objects}}
\startdata
\decimals
North ($p_{\rm CNN} > 0.0894$) & & \\
\phn\phn$19 < r < 21$ & $78.2$ & $0.911 \pm 0.067$ & $\sim 3.6\times 10^4$\\ % N = 0.911 * 8 * 5000
\phn\phn$21 < r < 21.15$ & $16.8$ & \nodata & \nodata \\
\phn\phn$r_{\rm fib} > 22$, $r < 21$ (in BGS) & $1.7$ & \nodata & \nodata \\
\hline
South ($p_{\rm CNN} > 0.1198$) & & \\
\phn\phn$19 < r < 21$ & $78.7$ & $0.867 \pm 0.056$ & $\sim 6.2 \times 10^4$ \\  % N = 0.867 * 8 * 9000
\phn\phn$21 < r < 21.15$ & 16.3 & \nodata & \nodata \\
\phn\phn$r_{\rm fib} > 22$, $r < 21$ (in BGS) & $1.6$ & \nodata & \nodata 
\enddata
\tablecomments{$p_{\rm CNN}$ refers to the threshold value for CNN-selected targets. Forecasts for completeness of the $z < 0.03$ sample and the total number of $z < 0.03$ galaxies are based on CNN cross-validation on $19 \leq r < 21$ objects and an assumed density of 8 low-$z$ objects per deg$^2$.}
\end{deluxetable*}

\subsection{CNN Retraining} \label{sec:cnn-retraining}

Our resulting LOW-Z sample provides a more comprehensive data set for retraining and validating the CNN.
During the same time period, the SAGA Survey has obtained more redshifts for objects within the projected virial radius of $z \sim 0.01$ host galaxies (Y.-Y. Mao, in preparation).
Our updated training set consists of 29,537 SAGA redshifts and 139,245 DESI LOW-Z objects within the expanded color--surface brightness selection that includes 95\% of $z < 0.03$ galaxies (Equations \ref{eq:g-r_fit} and \ref{eq:mu-r_fit}).
We retrain our CNN using the same architecture and framework described in Appendix~\ref{sec:cnn-details} using the updated redshift catalog.\footnote{We note that there are a small number of objects with high redshifts ($z > 0.3$) and $p_{\rm CNN} > 0.25$. Based on visual inspection, we believe that many of them are likely assigned incorrect redshifts, and we thus exclude them from our training set.} Because our aim is to form a complete survey of $z < 0.03$ galaxies and BGS has already demonstrated a high level of completeness for low-redshift objects, we evaluate the retrained CNN performance on targets outside of the BGS cuts.

We perform $k=5$-fold cross-validation and save each CNN model trained on an 80\% subset of the data; our results are based on the averaged predictions over the ensemble of 5 CNNs.
From the cross-validation results, we are able to assess the completeness as a function of $p_{\rm CNN}$, or equivalently, target density. A given $p_{\rm CNN}$ threshold corresponds to different number densities in the north and south skies due to differences in telescope instrumentation and BGS selection criteria.
We propose a CNN-selected target density of $95$ per deg$^2$ for objects outside of the BGS cuts and in the magnitude regime $19 \leq r < 21.15$, which corresponds to $p_{\rm CNN} > 0.0894$ in the northern sky and $p_{\rm CNN} > 0.1198$ in the southern sky.
For $19 \leq r < 21$ objects outside BGS cuts, we expect that the retrained CNN can achieve $85-90\%$ completeness for $z < 0.03$ objects.
This target list comprises $\sim 10^5$ objects at $r < 21$ that would otherwise not be targeted by BGS.
We also note that the $p_{\rm CNN}$ thresholds correspond to $>95\%$ completeness for objects in our entire redshift catalog (including BGS objects).
Our CNN forecasts for Y2 are shown in Table~\ref{tab:cnn_y2}.
We compare performance for the Y1 and Y2 CNNs in Appendix~\ref{sec:comparing-y1-y2-cnn}.

\subsection{LOW-Z Year 2 Selection}
\label{sec:Y2}

Our Year 2 sample, which began getting data in Fall 2022, consists of all objects between $19 \leq r < 21.15$ and $r_\mathrm{fib} < 23.5$ within the $z<0.03$-complete photometric cuts (Section \ref{sec:photo_cuts}) and excluding objects in the BGS Bright and Faint samples for objects with $r_{\rm fib} < 22$. The full sample of LOW-Z targets for Year 2 is then divided into two tiers of priority (Table \ref{tab:tier_sum}):

\textbf{Year 2 Tier 1} ($\sim 97$ objects per deg$^2$) consists of objects selected by the retrained CNN from the $z<0.03$-complete photometric cuts sample. Our selection includes the top-ranked $95$ objects per deg$^2$ outside BGS cuts in the $19 < r < 21.15$ range, in addition to the top $\sim 1.7$ CNN-selected objects per deg$^2$ with $r_{\rm fib} > 22$. This latter set of objects represents BGS targets that may encounter high redshift failure rates in bright observing time.  

\textbf{Year 2 Tier 2} ($\sim 325$ objects per deg$^2$) consists of all objects from the $z<0.03$-complete photometric cuts sample that are outside of the main BGS color cuts \citep{Ruiz_Macias_2020} or at $r_\mathrm{fib} > 22.0$ (and are not in Tier 1). 

We plan to continue to characterize the Y2 sample as new data comes in, although at present we do not plan to make further significant updates to our targeting strategy during the DESI main survey.

\subsection{Optimizing Catalog-Level Photometric Selection as a Function of Redshift for Future LOW-Z Surveys}
We parameterize the fit in Section \ref{sec:selection_eff} as a function of redshift and find that the redshift evolution of the intercept is well described by a linear fit. For the color cuts, the redshift evolution is described by the equation: 
\begin{equation}
    (g-r)_o - \sigma_{gr} + 0.06\,(r_o - 14) > 2.62 \times z + 0.90,
\label{eq:g-r_fit}
\end{equation}
while the surface brightness cuts evolve as: 
\begin{equation}
    \mu_{r{_o}, \textrm{eff}} + \sigma_{\mu} - 0.7 \, (r_o - 14)< -14.35 \times z + 17.33
\label{eq:mu-r_fit}
\end{equation}

This parameterization gives us a way to estimate target density as a function of redshift and magnitude for a complete LOW-Z survey using only catalog-level photometric information from DR9.  

An example of projected target density as a function of redshift for a range of cuts in apparent $r$-band magnitude is shown in Figure \ref{fig:density}. As expected, there is a trade-off between maximum apparent magnitude and redshift. We estimate that for $z < 0.03$, we could be complete out to $r < 21$ at 350 targets per square degree and $r < 22$ at 800 targets per square degree. We can use the GAMA luminosity and stellar mass functions to translate our completeness to a function of stellar mass \citep{Loveday150501003, Wright_2017}. For a survey of $z < 0.03$ galaxies out to $r < 22$ we expect to be complete for galaxies with M$_* > 10^7$ M$_\odot$. These results can help inform future planning for DESI-II and beyond on how to design an optimally targeted low-redshift survey.

\section{Conclusions}
We have described the DESI LOW-Z survey, a DESI secondary target program that has already generated a large and scientifically interesting survey of low-redshift objects and dwarf galaxies in the early stages of the DESI survey. This survey (including overlap with DESI BGS selection) includes over 140,000 objects with redshifts, over 22,000 dwarf galaxies ($M_* < 10^9 \text{M}_\odot$), and over 2,000 low-redshift objects ($z < 0.03$), rivaling SDSS and GAMA for the total number of low-redshift dwarf galaxies. Using the first few months of data from the DESI Y1 survey, we have validated the completeness of our photometric cuts at capturing the population of low-redshift galaxies. While we use all available low-redshift objects to evaluate our completeness, we note that the LOW-Z sample dominates the data set. We have also studied the properties of a CNN-selected sample with lower target density, trained on low redshift data from the SAGA survey.

We find that:
\begin{enumerate}
 \item Our $z<0.03$-complete photometric cuts are $\sim 95\%$ complete at $z < 0.03$ between $19 < r < 21$.
    \item  Our CNN is approximately 20\% efficient at selecting low-redshift galaxies, compared to efficiencies of $\sim 1\%$ using traditional photometric methods. 
    \item We achieve $\sim 75\%$ fiber allocation for objects that overlap with BGS and $\sim 30\%$ fiber allocation for objects outside of the BGS Bright and BGS Faint samples for a combined fiber allocation fraction of $\sim 50\%$.
    \item We find no evidence of increasing redshift failures with $r$-band magnitude, but see a strong increase in the redshift failure rate as a function of $r_\mathrm{fib}$ for objects at $r_\mathrm{fib} > 23$ in dark time and $r_\mathrm{fib} > 22$ in bright time. We also find that this increase in redshift failure is correlated with $p_{\rm CNN}$ at fixed $r_\mathrm{fib}$, indicating somewhat lower redshift success for true low-redshift galaxies.
    \item The LOW-Z survey is currently observing $3.7$ low-redshift galaxies ($z < 0.03$) per square degree. We expect this to be a lower limit for DESI Y2 observations, given improved targeting strategies. 
\end{enumerate}
Based on these data, we have retrained a new CNN to select a complete and efficient sample of low-redshift galaxies. Using this retrained CNN, we estimate that we can achieve $85-90\%$ completeness within our catalog-level photometric cuts to $z < 0.03$ with $\sim 80$ targets per square degree for $19 < r < 21$. Using this information, we update our Y2 targeting strategy to target objects outside of the BGS survey to a slightly fainter magnitude limit ($r < 21.15$) with a fiber magnitude cut at $r_{\rm fib} < 23.5$. 

Beyond Y2, the LOW-Z survey provides a blueprint for the design of a higher-priority low-redshift survey as part of Y3--Y5 or DESI II. In the future, we estimate that we could run a complete low-redshift survey ($z < 0.03$) at 350 targets per square degree at $z < 21$ or 800 targets per square degree at $z < 22$. Translating this to stellar mass would correspond to a complete survey for galaxies with $M_* > 10^{7.5} \text{M}_\odot$ or $M_* > 10^{7.0} \text{M}_\odot$ respectively. Such a dense map of the local universe would provide an incredibly rich dataset for studying the local density and velocity field and the relation of galaxy properties to this field, for identifying the host galaxies of transients and gravitational waves, and for expanding our understanding of galaxy formation at the lowest masses.  

\section*{Data Availability}
LOW-Z data will be released as part of the DESI data releases. A portion of the data analyzed here will be released as part of the Early DESI Data Release, expected in mid 2023.

All data points shown on the figures are available in a machine-readable form at \dataset[DOI: 10.5281/zenodo.7422591]{https://zenodo.org/record/7422591}
%\url{https://zenodo.org/record/7422591}. 

\begin{acknowledgments}
We thank Mia de los Reyes and Kelly Douglass for their helpful comments on the draft. We would also like to thank the DESI collaboration internal reviewers, Rita Tojeiro and Jeremy Tinker, for helpful feedback that improved the paper. We also thank Mike Blanton for suggesting an observational bias that could explain lower redshift success rates for lower-redshift galaxies.  We are grateful to the anonymous referee for comprehensive comments that significantly improved the presentation of the paper.

This work received support from the Kavli Institute for Particle Astrophysics and Cosmology at Stanford University and SLAC National Accelerator Laboratory and from the U.S. Department of Energy under contract number DE-AC02-76SF00515 to SLAC National Accelerator Laboratory. Support for YYM was partly provided by NASA through the NASA Hubble Fellowship grant no.\ HST-HF2-51441.001, awarded by the Space Telescope Science Institute, which is operated by the Association of Universities for Research in Astronomy, Incorporated, under NASA contract NAS5-26555. 

The SAGA Survey (sagasurvey.org) is a spectroscopic survey with data obtained from the Anglo-Australian Telescope, the MMT Observatory, and the Hale Telescope at Palomar Observatory. The SAGA Survey made use of public imaging data from the Sloan Digital Sky Survey (SDSS), the DESI Legacy Imaging Surveys, and the Dark Energy Survey, and also public redshift catalogs from SDSS, GAMA, WiggleZ, 2dF, OzDES, 6dF, 2dFLenS, and LCRS. The SAGA Survey was supported by NSF collaborative grants AST-1517148 and AST-1517422 to RHW and MG and by Heising–Simons Foundation grant 2019-1402.

The DESI project is supported by the Director, Office of Science, Office of High Energy Physics of the U.S. Department of Energy under Contract No.DE-AC02-05CH11231, and by the National Energy Research Scientific Computing Center, a DOE Office of Science User Facility under the same contract; additional support for DESI is provided by the U.S. National Science Foundation, Division of Astronomical Sciences under Contract No. AST-0950945 to the NSF’s National Optical-Infrared Astronomy Research Laboratory; the Science and Technologies Facilities Council of the United Kingdom; the Gordon and Betty Moore Foundation; the Heising-Simons Foundation; the French Alternative Energies and Atomic Energy Commission (CEA); the National Council of Science and Technology of Mexico; the Ministry of Economy of Spain, and by the DESI Member Institutions. The authors are honored to be permitted to conduct astronomical research on Iolkam Du’ag (Kitt Peak), a mountain with particular significance to the Tohono O’odham Nation.
\end{acknowledgments}

\bibliography{main}
\bibliographystyle{aasjournal}

\appendix
\section{CNN Optimization Details} \label{sec:cnn-details}

We train CNNs to identify $z < 0.03$ galaxies from optical image cutouts. The optimized model acts as a mapping between images ($x \in \mathbb{R}^{3 \times 144 \times 144}$) to a scalar prediction, $p_{\rm CNN} \in [0, 1]$; if $p_{\rm CNN}$ exceeds some threshold value, then the input image can be classified as a low-redshift galaxy candidate.
Because modern neural networks have $\mathcal O(10^7)$ tunable parameters \citep{He2016}, the optimization process must be done carefully.
We closely follow the methodology of \citet{Wu2021xsaga}, which uses a trained CNN to identify $z < 0.03$ galaxies with balanced purity and completeness (i.e., similar levels of false positives and false negatives).
In this work, we have selected galaxies at a higher level of completeness at the cost of lower purity (accomplished by using a threshold for $p_{\rm CNN}$ below the value of $0.5$ used by \citealt{Wu2021xsaga}).

We use an extended version of the \cite{mao2020saga} SAGA redshift catalog as the ground truth data set for training the model. We use an 80\%/20\% training/validation split, such that a random 80\% subset is used for training, and the remaining 20\% is used for evaluating the CNN. 
We focus on the accuracy, purity (precision), and completeness (recall) metrics for evaluating CNN performance on $z < 0.03$ predictions.
These metrics allow us to compare different combinations of hyperparameters, such as the model architecture, optimization objective, and optimization schedule.
By examining these validation metrics, we can gauge whether the model has ``overfit,'' leading to strong performance on the training data but poor generalization on unseen data, or if the model has converged.

We briefly summarize the hyperparameter choices used in our CNN model.
Our model architecture is a 34-layer residual neural network with several modifications that allow for efficient processing of sparse astronomical images \citep{WuPeek2020}.
Because the training data are heavily imbalanced in favor of high-$z$ examples, we adopt the Focal Loss function for optimization \citep{Lin2017}.
We train the CNN using the Ranger optimizer\footnote{\url{https://github.com/lessw2020/Ranger-Deep-Learning-Optimizer}} and a one-cycle schedule for the learning rate and momentum hyperparameters \citep{Smith2018} for ten epochs.

\section{Comparison of the Y1 and Y2 trained CNNs}
\label{sec:comparing-y1-y2-cnn}

\begin{figure*}[t]
    \centering
    \includegraphics[width=\columnwidth]{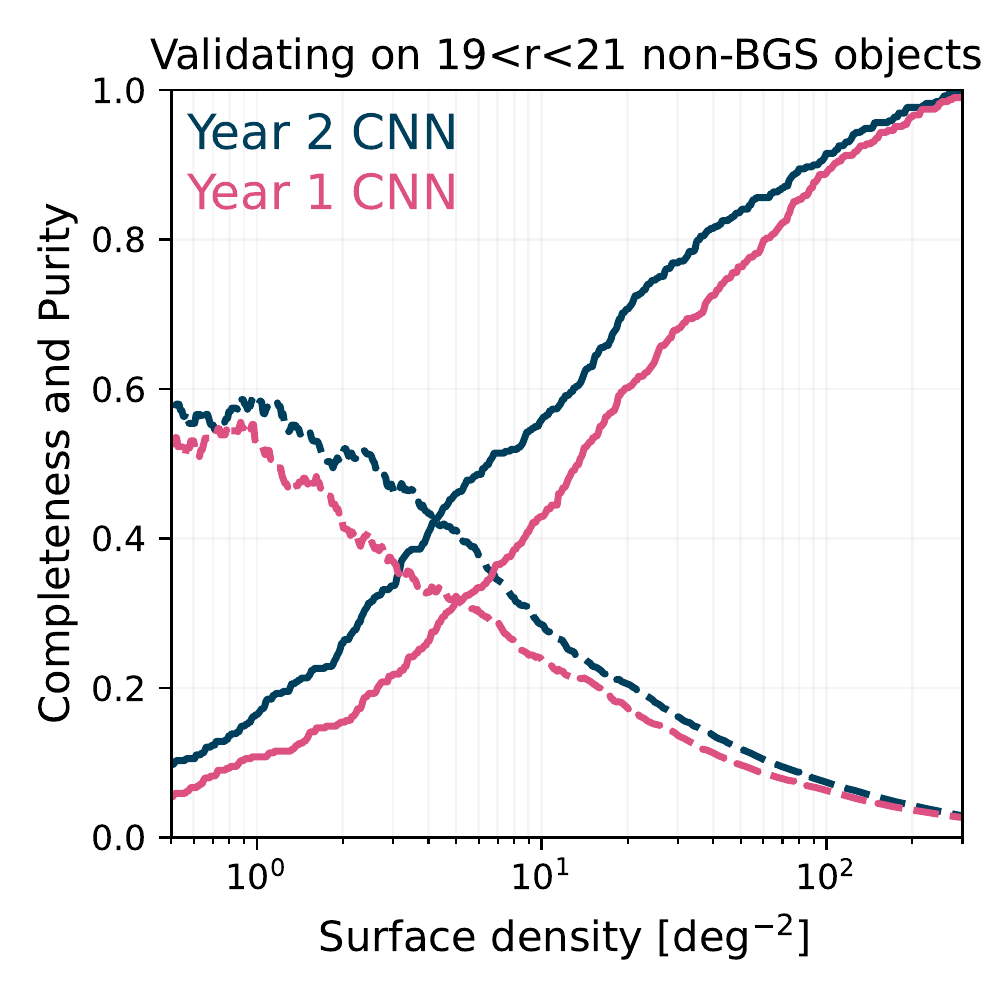}
    \includegraphics[width=\columnwidth]{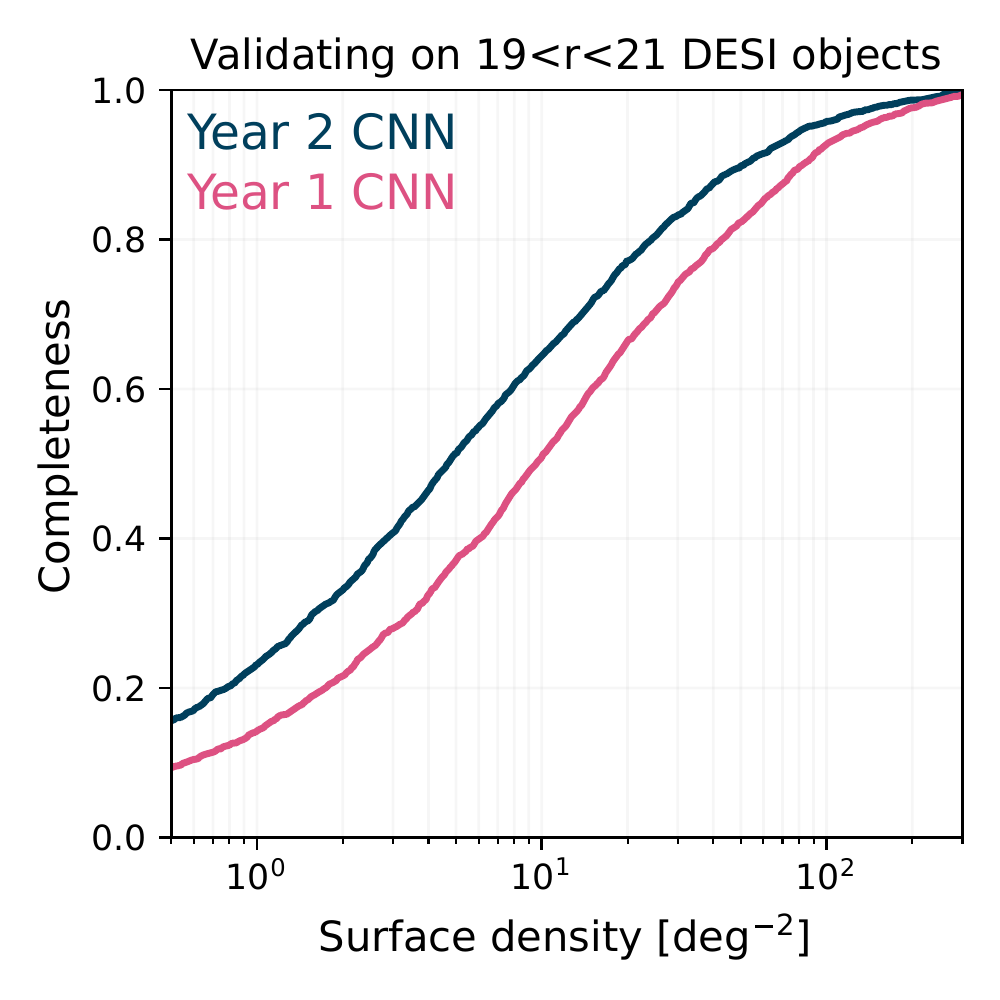}
    \caption{Estimates of the CNN selection completeness (\textit{solid}) and purity (\textit{dashed}) for low-redshift objects as a function of target density for non-BGS objects (\textit{left}) and all DESI objects (\textit{right}) in the magnitude range $19 < r < 21$. We find that the Y2 retrained CNN (\textit{blue}) outperforms the Y1 CNN (\textit{magenta}) in terms of both purity and completeness.}
    \label{fig:CNN-retraining-performance}
\end{figure*}

The Y1 CNN and the Y2 retrained CNN are validated on $19 < r < 21$ objects with redshifts in the DESI survey.  
In Figure~\ref{fig:CNN-retraining-performance}, we estimate the low-redshift purity and completeness as a function of target density.
Objects that fall outside BGS cuts are shown in the left panel, while all objects in DESI are shown in the right panel of Figure~\ref{fig:CNN-retraining-performance}.

The Y2 CNN purity and completeness are determined using $k=5$ cross-validation in order to ensure independent training/validation sets, while the Y1 CNN performance is characterized using a single CNN trained on all of the then-available data; this may result in a slight underprediction of the Y2 performance.
Additionally, the Y1 CNN was used to select part of the sample that was used for cross-validation, thereby inflating the Y1 CNN completeness in that regime.
Nonetheless, we find that the Y2 retrained CNN has improved completeness and purity as a result of its larger training set (see Section~\ref{sec:cnn-retraining}).

\end{document}